\documentclass[prd,aps,floats,floatfix,eqsecnum,nofootinbib]{revtex4}
\usepackage{amsmath,amssymb,verbatim,epsfig,graphicx,rotating}
\usepackage{psfrag}
\newcommand{\be}{\begin{equation}}
\newcommand{\ee}{\end{equation}}
\newcommand{\bea}{\begin{eqnarray}}
\newcommand{\eea}{\end{eqnarray}}
\begin{document}
\title{Single Field Inflation models allowed and ruled out by
the three years WMAP data}
\author{\bf H. J. de Vega$^{(b,a)}$}\email{devega@lpthe.jussieu.fr}
\author{\bf N. G. Sanchez $^{(a)}$}\email{Norma.Sanchez@obspm.fr}
\affiliation{$^{(a)}$
Observatoire de Paris, LERMA, Laboratoire Associ\'e au CNRS UMR 8112,
 \\61, Avenue de l'Observatoire, 75014 Paris, France. 
\\$^{(b)}$ LPTHE, Laboratoire Associ\'e au CNRS UMR 7589,\\
Universit\'e Pierre et Marie Curie (Paris VI) et Denis Diderot (Paris VII),\\
Tour 24, 5 \`eme. \'etage, 4, Place Jussieu, 75252 Paris, Cedex 05,
France.}

\begin{abstract}
We study the single field slow-roll inflation models  that
better agree with the available CMB and LSS data including the three years WMAP data:
new inflation and hybrid inflation. We study these models as effective field 
theories in the Ginsburg-Landau context: a trinomial potential turns out to 
be a simple and well motivated model.
The spectral index $ n_s $ of the adiabatic fluctuations, the ratio $ r $
of tensor to scalar fluctuations and the running  index $ d n_s/d 
\ln k $ are studied in detail.  We derive explicit formulae for 
$ n_s , \; r $ and $ d n_s/d \ln k $ and provide relevant plots. 
In new inflation, and for the three years WMAP and 2dF central value $ n_s = 0.95 $,
we predict $ 0.03 < r < 0.04 $ and $ -0.00070 < d n_s/d \ln k < -0.00055 $. 
In hybrid inflation, and for $ n_s = 0.95 $,  we predict $ r \simeq 0.2 $ and 
$ d n_s/d \ln k \simeq -0.001 $ . Interestingly enough, 
we find that in new inflation $ n_s $ is {\bf bounded} from above by 
$ n_{s~max} = 0.961528 \ldots $ and that $ r $ is a {\bf two} valued function of $ n_s $ 
in the interval $ 0.96 < n_s <  n_{s~max} $. In the first branch 
we find $ r < r_{max} = 0.114769\ldots $. In hybrid inflation we find a 
critical value $ \mu_{0~crit}^2 $ for the mass parameter $ \mu_0^2 $ of the 
field $ \sigma $ coupled to the inflaton. 
For $ \mu_0^2 < \Lambda_0 \; M_{Pl}^2/192 $, where  $ \Lambda_0 $ is the 
cosmological constant, hybrid inflation is ruled out by the WMAP three years
data since it yields a blue tilted $ n_s > 1 $ behaviour. 
Hybrid inflation for $ \mu_0^2 > \Lambda_0 \; M_{Pl}^2/192 $ fullfills all 
the present CMB+LSS data for a large enough initial inflaton amplitude. 
Even if chaotic inflation predicts $ n_s $ values 
compatible with the data,  chaotic inflation is disfavoured since it predicts 
a too high value $ r \simeq 0.27 $ for the ratio of tensor to scalar 
fluctuations. The model which  best agrees with the current data and which best 
prepares the way to the expected data $ r \lesssim 0.1 $, is the trinomial 
potential with negative mass term: new inflation.
\end{abstract}

\date{\today}
\maketitle
\tableofcontents

\section{Introduction and Results}

Inflation was  introduced  to solve several outstanding
problems of the standard Big Bang model \cite{guth} and became an important
part of the standard cosmology. At the same time, it provides a natural
mechanism for the generation of scalar density fluctuations that
seed large scale structure, thus explaining the origin of the
temperature anisotropies in the cosmic microwave background (CMB),
as well as that of  tensor perturbations (primordial gravitational
waves)\cite{mukyotr,libros}. 

\medskip

A distinct aspect of
inflationary perturbations is that these are generated by quantum
fluctuations of the scalar field(s) that drive inflation. After
their wavelength becomes larger than the Hubble radius, these
fluctuations are amplified and grow, becoming classical and
decoupling from  causal microphysical processes. Upon re-entering
the horizon, during the matter era, these classical perturbations
seed the inhomogeneities which generate structure upon
gravitational collapse\cite{mukyotr,libros}. A great
diversity of inflationary models predict fairly generic features:
a gaussian, nearly scale invariant spectrum of (mostly) adiabatic
scalar and tensor primordial fluctuations, making the inflationary
paradigm fairly robust. The gaussian, adiabatic and nearly scale
invariant spectrum of primordial fluctuations provide an excellent
fit to the highly precise wealth of data provided by the Wilkinson
Microwave Anisotropy Probe (WMAP)\cite{WMAP,WMAP3}
Perhaps the most striking validation of inflation as a mechanism for generating
\emph{superhorizon} (`acausal')  fluctuations is the
anticorrelation peak in the temperature-polarization (TE) angular
power spectrum at $l \sim 150$ corresponding to superhorizon
scales\cite{WMAP}. The confirmation of many of the robust predictions of 
inflation by current high precision observations places
inflationary cosmology on solid grounds.

\medskip

Amongst the wide variety of inflationary scenarios, single field
slow roll models provide an
appealing, simple and fairly generic description of inflation. Its
simplest implementation is based on a scalar field (the inflaton)
whose homogeneous expectation value drives the dynamics of the
scale factor, plus small quantum fluctuations. The inflaton
potential, is fairly flat during inflation. This flatness not only
leads to a slowly varying Hubble parameter, hence ensuring a
sufficient number of e-folds, but also provides an explanation for
the gaussianity of the fluctuations as well as for the (almost)
scale invariance of their power spectrum. A flat potential
precludes large non-linearities in the dynamics of the
\emph{fluctuations} of the scalar field. 

\medskip

The current WMAP data
seem to validate the simpler one-field slow roll
scenario \cite{WMAP,WMAP3}. Furthermore, because the potential is flat
the scalar field is almost {\bf massless}, and modes cross the horizon
with an amplitude proportional to the Hubble parameter. This fact
combined with a slowly varying Hubble parameter yields an almost
scale invariant primordial power spectrum.  The slow-roll approximation has
been recently cast as a $1/N_{efolds}$ expansion \cite{1sN}, where
$ N_{efolds} \sim 50 $ is  the number of efolds before the end of inflation
when modes of cosmological relevance today first crossed the Hubble
radius.

The observational progress permit to start 
to discriminate among different inflationary models, placing
stringent constraints on them. The upper bound on the ratio $ r $ 
of tensor to scalar fluctuations obtained by WMAP \cite{WMAP,WMAP3} 
rules out the massless $ \phi^4 $ model and {\bf necessarily} implies 
the presence of a {\bf mass term} in the inflaton potential \cite{1sN,WMAP3}.

Besides its simplicity, the trinomial potential is a physically well motivated
potential for inflation in the grounds of the Ginsburg-Landau approach
to effective field theories (see for example ref.\cite{quir}).
This potential is rich enough to describe the physics of inflation and 
accurately reproduce the WMAP data \cite{WMAP,WMAP3}. 

The slow-roll expansion plus the WMAP data constraints the inflaton potential
to have the form \cite{1sN}
\be \label{Vi} 
V(\phi) = N_{efolds} \; M^4 \; w(\chi) \; ,
\ee  
\noindent where $ \phi $ is the inflaton field,
$\chi$ is a dimensionless, slowly varying field 
\be\label{chiflai} 
\chi \equiv \frac{\phi}{\sqrt{N_{efolds}} \;  M_{Pl}}  \; ,
\ee 
\noindent $ w(\chi) \sim \mathcal{O}(1) $ and 
$ M $ is the energy scale of inflation which is determined by the 
amplitude of the scalar adiabatic fluctuations \cite{WMAP} to be
$$ 
M \sim 0.00319 \; M_{Pl} = 0.77 \times 10^{16} {\rm GeV} \; .
$$
Following the spirit of the Ginsburg-Landau theory of phase transitions,
the simplest choice is a quartic trinomial for the inflaton potential
\cite{nos,1sN}:
\be \label{wxi}
w(\chi)= w_0 \pm \frac12 \; \chi^2 + \frac{h}3 \; \sqrt{\frac{y}2} \; \chi^3 +
\frac{y}{32} \; \chi^4 \; .
\ee
where the coefficients $ w_0, \; h $ and $ y $ are dimensionless and of order 
one and the signs $ \pm $ correspond to large and small field inflation, 
respectively (chaotic and new inflation, respectively).
Inserting eq.(\ref{wxi}) in eq.(\ref{Vi}) yields,
\be\label{VI}
V(\phi)= V_0 \pm \frac{m^2}{2} \; \phi^2 +  \frac{ m
\; g }{3} \; \phi^3 + \frac{\lambda}{4}\; \phi^4 \; .
\ee
where the mass $ m^2 $ and the couplings $ \; g $ and $ \lambda $ are given
by the following see-saw-like relations, 
\be 
m = \frac{M^2}{M_{Pl}} \qquad ,  \qquad g = h \; \sqrt{\frac{y}{2 \; N}} 
\; \left( \frac{M}{M_{Pl}}\right)^2  \qquad ,  \qquad \lambda  = 
\frac{y}{8 \; N} \left( \frac{M}{M_{Pl}}\right)^4 \label{acoi} 
\qquad ,  \qquad  V_0 = N \; M^4 \; w_0 \; . 
\ee 
where $ N \equiv N_{efolds}$. 
Notice that $ y \sim {\cal O}(1) \sim h $ guarantee that $ g  \sim 
{\cal O}(10^{-6}) $ and $ \lambda  \sim {\cal O}(10^{-12}) $ without
any fine tuning as stressed in ref. \cite{1sN}. That is, the smallness
of the couplings  directly follow from the form of the inflaton potential
eq.(\ref{Vi}) and the amplitude of the scalar fluctuations that fixes $ M $
\cite{1sN}.

\medskip

The small coupling limit $ y \to 0 $ of eqs.(\ref{wxi})-(\ref{VI}) corresponds 
to a quadratic potential while the strong coupling limit $ y \to \infty $
yields the massless quartic potential. The extreme asymmetric limit
$ |h| \to \infty $ yields a massive model without quadratic term. In such limit
the product $ |h| \; M^2 $ must be kept fixed since it is determined by the 
amplitude of the scalar fluctuations.

\medskip

We study here new inflation with the trinomial potential 
eqs.(\ref{wxi})-(\ref{VI}) and hybrid inflation [see below], the two models
fulfill the observational constraints. We compute in both scenarios 
$ n_s , \; r $ and the running $ d n_s/d \ln k $ as functions of the 
parameters of the models, derive explicit formulae for 
$ n_s , \; r $ and $ d n_s/d \ln k $ and provide relevant plots. 
Moreover, we plot the ratio $ r $ and the running $ d n_s/d \ln k $ 
as functions of the scalar index $ n_s $. Since the value of $ n_s $ is now
known \cite{WMAP3}-\cite{Teg}, these plots allow us to {\bf predict} the values of 
$ r $ and $ d n_s/d \ln k $ for the different inflationary models considered.
These predictions and plots are solely produced from theory and not from any 
fitting of the data.

\bigskip

The three years WMAP data indicate a red tilted spectrum ($ n_s < 1 $)
with a small ratio $ r < 0.28 $  of tensor to scalar fluctuations \cite{WMAP3}.
The present data do not permit to find the precise values neither
of the ratio $ r $ nor of the running index $ d n_s/d \ln k $, 
only upper bounds are obtained \cite{WMAP,WMAP3}. We therefore think
that the value of $ n_s $ [eq.(\ref{nswmap})] obtained through a fit of the 
data assuming $ r =  d n_s/d \ln k = 0 $ is more precise than the values of 
$ n_s $ obtained through fits allowing both $ r $ and $ d n_s/d \ln k $ to vary.
Notice that $ n_s = 0.95 $ was independently found from the 2dF data
under similar assumptions \cite{2dF}. More precisely, from the three years WMAP 
data \cite{WMAP3} as well as ref. \cite{2dF} we take
\be \label{nswmapi}
 n_s = 0.95 \pm 0.02 \; .
\ee
We find that for $ n_s = 0.95 $ and any value of the asymmetry $ h $ 
[see figs. \ref{nsr} and \ref{nsrun}], new inflation with the trinomial potential 
eqs.(\ref{wxi})-(\ref{VI}) predicts
$$
{\rm trinomial~potential~new~inflation~for~} n_s = 0.95 : \quad 
0.03 < r < 0.04 \quad {\rm and}\quad -0.00070 < d n_s/d \ln k < -0.00055 \; .
$$
We find for the lower value $ n_s = 0.93 $ of the three years WMAP data band:
$$
{\rm trinomial~potential~new~inflation~for~} n_s = 0.93 : \quad 
0.003 < r < 0.015 \quad {\rm and}\quad -0.0011 < d n_s/d \ln k < -0.00033 \; .
$$
Moreover, in new inflation with the trinomial potential, we find that $ n_s $ 
is {\bf bounded} from above by  
$$ 
{\rm new~inflation:} \qquad n_s < n_{s~maximum} = 0.961528 \ldots \; .
$$ 
For $ n_s = 0.961528 \ldots $ we have in this model $ r =  0.114769\ldots $
(see figs. \ref{nsr} and \ref{nsh}). 
Interestingly enough, there exists {\bf two} values (two branches) of
$ r $ for one value of  $ n_s $ in the interval $ 0.96 < n_s < 0.961528 
\ldots $ [see fig. \ref{nsr}]. The value $ r_{max} = 0.114769\ldots $ is the 
maximun $ r $ in the first branch. The values $ 0.16 \geq r \geq 
0.114769\ldots $ correspond to a second branch of $ r $ as a function of 
$ n_s $ in the interval $ 0.96 < n_s < 0.961528 \ldots $. 
In the first branch we have 
$$
 r_{max} = 0.114769\ldots \; .
$$
The absolute maximun value $ r_{abs~max} = 0.16 $ belongs to the second branch
and corresponds to the quadratic monomial potential obtained from eq.(\ref{wxi}) 
at $ y = 0 $.

These predicted values of the ratio $ r $ fullfil the three years WMAP bound
including SDSS galaxy survey \cite{WMAP3}
\be \label{rwmapi}
r < 0.28 \; (95\% ~CL) \; .
\ee
Moreover, one can see from fig. 14 in ref. \cite{WMAP3} 
that $ r < 0.1 ~(68\% ~CL) $ from WMAP$+$SDSS. 

\medskip

Chaotic inflation  with the trinomial potential eq.(\ref{wxi})-(\ref{VI})
yields larger values of $ r $ than new inflation
for a given value of $ n_s $ \cite{nos}. More precisely, for $ n_s = 0.95 $
we find $ r = 0.27 $ for the binomial potential \cite{nos}
(the trinomial potential introduces very small changes).

Therefore, although the WMAP value for $ n_s $ [eq.(\ref{nswmapi})]
is compatible both with chaotic and new inflation, the WMAP bounds on 
$ r $ {\bf clearly disfavour} chaotic inflation. New inflation easily fulfils 
the three years WMAP bounds on $ r $ and prepares the way for the expected
 data on the ratio of tensor/scalar fluctuations $ r \lesssim 0.1 $.

\bigskip

In the inflationary models of hybrid type, the inflaton is coupled to another 
scalar field $ \sigma_0 $ with mass term $ -\mu_0^2 < 0 $
through a potential of the type \cite{lin}
\bea\label{Vhib1i}
&&V_{hyb}(\phi,\sigma_0) = \frac{m^2}{2} \; \phi^2 + \frac{g_0^2}{2} \; 
\phi^2 \; \sigma_0^2 + \frac{\mu_0^4}{16 \, \Lambda_0} 
\left(\sigma_0^2 -  \frac{4 \, \Lambda_0}{\mu_0^2} \right)^2= \cr \cr
&&  =\frac{m^2}{2} \; \phi^2 + \Lambda_0+\frac12 \; (g_0^2 \; 
\phi^2-\mu_0^2) \;  \sigma_0^2+\frac{\mu_0^4}{16 \; \Lambda_0} \; 
\sigma_0^4\; ,
\eea
where $ m^2 > 0 , \; \Lambda_0 > 0 $ plays the role of a cosmological constant 
and $ g_0^2 $  couples $ \sigma_0 $ with $ \phi $.

\medskip

The initial conditions are chosen such that $\sigma_0$ and $ \dot\sigma_0$ 
are very small (but not identically zero) and therefore 
inflation is driven by the cosmological constant $ \Lambda_0 $ 
plus the initial value of the inflaton $  \phi(0) $.
The inflaton field $ \phi(t) $ decreases with time while the scale factor 
$ a(t) $ grows exponentially with time. The  field $ \sigma_0 $ has an
effective classical mass square
\be\label{masefei}
m_{\sigma}^2 = g_0^2 \; \phi^2 - \mu_0^2 \; .
\ee
Since  the inflaton field $ \phi $ decreases with time, $ m_{\sigma}^2 $ 
becomes 
negative at some moment during inflation. At such moment, spinodal (tachyonic) 
unstabilities appear and the field $ \sigma_0 $ starts to grow exponentially. 
Inflation stops when both fields $ \phi $ and $ \sigma_0 $ are comparable with 
$ \dot \phi $ and  $ \dot \sigma_0 $ and close to their vaccum values.

We find that the time when the effective mass of the field $ \sigma_0 $ 
eq.(\ref{masefei}) becomes negative depends on the
values of $ \mu_0^2 $ and $ g_0^2 \; \phi^2(0) $. For low values of 
$ \mu_0^2 $ the field  $ \sigma_0 $ starts to grow
close to the end of inflation. On the contrary, for higher values of 
$ \mu_0^2 $ the field  $ \sigma_0 $ starts to grow
well before the end of inflation.  This is explained by the fact
that the scale of time variation of $ \sigma_0 $ goes as
$ \mu_0^{-1} $.  $ \sigma_0 $ evolves slowly for small $ \mu_0 $ and
fastly for large $ \mu_0 $ [see figs. \ref{a}-\ref{pe}].

{\bf Only} at $ \Lambda = 0 $ hybrid inflation becomes chaotic inflation 
with the monomial potential $ \frac{m^2}{2} \; \phi^2 $. For any value of 
$ \Lambda > 0 $ even very small, the features of hybrid inflation remain.

We compute $ n_s , \; r $ and  $ d n_s/d \ln k $ for hybrid inflation
as functions of the parameters in the potential eq.(\ref{Vhib1i}) and
the initial value of the inflaton field 
[see figs. \ref{nsr005A}-\ref{nsrunA}]. 

The results of our extended numerical investigation of hybrid inflation can
be better expressed in terms of the dimensionless variables
$$
\Lambda \equiv \frac{2 \, \Lambda_0}{M^4 \; N_{efolds}} \quad , \quad
{\hat \chi} \equiv \frac{\phi}{\sqrt{N_{efolds} \; \Lambda} \;  M_{Pl}} 
\quad {\rm and} \quad {\hat \mu}^2 \equiv \frac{\mu_0^2 \;  
M_{Pl}^2 \; N_{efolds}}{2 \; \Lambda_0} \; .
$$
We depict in figs. \ref{nsr005A}-\ref{nsrunA} the observables 
$ n_s , \; r $ and the running index 
$ d n_s /d \ln k $ as  functions of $ \Lambda $ and $ n_s $.
We present a complete picture for hybrid inflation covering {\bf two} 
different, blue tilted and red tilted, regimes. We find that for all the 
observables, the shape of the curves depends crucially on the mass parameter 
$ {\hat \mu}^2 $ of the $\sigma$ field and the (rescaled) initial amplitude 
$ {\hat \chi}(0) $ of the inflaton field.

We find a blue tilted spectrum ($ n_s > 1 $) for 
$ {\hat \mu}^2 <  {\hat \mu}^2_{crit}
\simeq 0.13 $ while for   $ {\hat \mu}^2 >  {\hat \mu}^2_{crit} $ we can have
either $ n_s > 1 $ or  $ n_s < 1 $ depending on the initial conditions:
for $ {\hat \chi}(0) > {\hat \chi}(0)_{crit} $ we have $ n_s > 1 $,
and for $ {\hat \chi}(0) < {\hat \chi}(0)_{crit} $ we have $ n_s < 1 $. 
The value of $ {\hat \chi}(0)_{crit} $ grows with $ {\hat \mu}^2 $:
for $ {\hat \mu}^2 = 0.5 $,  we find  $ {\hat \chi}(0)_{crit} = 2.7 $
and for $ {\hat \mu}^2 = 1.7 $, we find   $ {\hat \chi}(0)_{crit} = 5.8 $.

We see that $ n_s > 1 $ happens when the cosmological constant 
$ \Lambda_0 $ is large enough compared with $  \mu_0^2 \;  M_{Pl}^2 \; 
N_{efolds}$. More precisely,  $ {\hat \mu}^2 <  {\hat \mu}^2_{crit} $ for
$ \Lambda_0 > 192 \; \mu_0^2 \;  M_{Pl}^2 $ using $ N_{efolds} = 50 $.
That is, for $ \Lambda_0 < 192 \; \mu_0^2 \;  M_{Pl}^2$ we have either
red or blue tilted spectrum as explained above.

\medskip

For large $\Lambda, \;   n_s -1, \; r $ and $ d n_s/d \ln k $ always tend 
asymptotically to zero whatever be $ {\hat \mu}^2 $ and $ {\hat \chi}(0) $.

\medskip

We see from our calculations that all blue tilted values of  
$ (n_s, r) $ in the domain $ 1 < n_s < 1.15 , \; 0 < r < 0.2 $ 
can be realized by the hybrid inflation model eq.(\ref{Vhib1i}).
However, at the light of the three years WMAP data ref. \cite{WMAP3}
the blue tilted regime in hybrid inflation $ \mu^2 <  \mu^2_{crit} $ 
is ruled out.

The situation is totally different in the red tilted regime $  {\hat  \mu}^2 >
{\hat \mu}^2_{crit} \simeq 0.13 $ in hybrid inflation. 
The possible values of $ (n_s, r) $ for such regime of hybrid inflation are 
in the upper-right quadrant as shown in fig. \ref{bordenuevo}.

Hybrid inflation in the red tilted regime $ \mu^2 >  \mu^2_{crit} $
and $ {\hat \chi}(0) <  {\hat \chi(0)}_{crit} $ 
fulfills the three years WMAP value for $ n_s $ [see eq.(\ref{nswmap})] as
well as the bound on the ratio $ r $ [eq.(\ref{rwmap})].
We can read from fig. \ref{nsrY} and \ref{nsrunY} that 
$$ 
\mu^2 >  \mu^2_{crit} \; \; {\rm hybrid ~ inflation:} \quad 
0.2 > r > 0.14 \quad {\rm and } \quad -0.001 < d n_s /d \ln k < 0 \quad
{\rm for} \quad  0.952 <  n_s < 0.97  \; .
$$
Notice that hybrid inflation in the red tilted regime yields a too large
ratio $ r > 0.2 $ for $ n_s < 0.95 $.

\medskip

At the central three years WMAP value  $ n_s = 0.95 $ both new and hybrid 
inflation
are allowed. However, for $ n_s < 0.95 $ hybrid inflation is in trouble 
($ r > 0.2 $)
while for  $ n_s > 0.962 $ new inflation is excluded.

\medskip

The potential which best agree with the present red tilted spectrum 
and which best prepares the way to the expected data 
(a small $ r \lesssim 0.1 $) is the trinomial  potential eqs.(\ref{wxi})-(\ref{VI}) 
with negative mass term, that is 
small field (new) inflation. Hybrid inflation with a trinomial potential
can also reproduce the present data in the red tilted regime 
$ \mu^2 >  \mu^2_{crit} $ and $ {\hat \chi}(0) <  {\hat \chi(0)}_{crit} $.

\medskip

All calculations presented in this paper stem from the inflaton potential
in the slow roll approximation (dominant order in $ 1/N \simeq 1/50 $).
They do not use observational data as input. The analytical formulas 
  and plots provided in the paper allow to read directly the predicted
  values of $ r $ and $ dn_s/d \ln k $ as functions of $ n_s $.  In order to make illustrative
  predictions, we take the value $ n_s = 0.95 \pm 0.02$, as a judicious
  choice. The reader can see directly from the plots presented here
 our predictions for $ r $ and $ d n_s /d \ln k $ for
 future observational values of $ n_s $.

\section{The Inflaton Potential and the $ 1/N_{efolds} $ Slow Roll Expansion}

The description of cosmological inflation is based on an isotropic
and homogeneous geometry, which assuming flat spatial sections is
determined by the invariant distance
\be
ds^2= dt^2-a^2(t) \; d\vec{x}^2 \label{FRW} \; .
\ee
The scale factor obeys the Friedman equation
\be \label{ef}
\left[ \frac{1}{a(t)} \; \frac{da}{d t} \right]^2 =
\frac{\rho( t)}{3 \; M^2_{Pl}}   \;  ,
\ee
where $M_{Pl}= 1/\sqrt{8\pi G} = 2.4\times 10^{18}\,\textrm{GeV}$.

In single field inflation the energy density is dominated by a
homogeneous scalar \emph{condensate}, the inflaton, whose dynamics
is described by an  \emph{effective} Lagrangian
\be\label{lagra}
{\cal L} = a^3({ t}) \left[ \frac{{\dot
\phi}^2}{2} - \frac{({\nabla \phi})^2}{2 \;  a^2({ t})} - V(\phi) \right]
\; .
\ee
\noindent The inflaton potential $ V(\phi) $ is a slowly varying
function of $ \phi $ in order to permit a slow-roll solution for
the inflaton field $ \phi(t) $.

 We showed in ref. \cite{1sN} that combining the WMAP data with the
 slow roll expansion yields an inflaton potential of the form 
\be \label{V} 
V(\phi) = N \; M^4 \; w(\chi)  \; ,
\ee  
\noindent where $\chi$ is a dimensionless, slowly varying field 
\be\label{chifla} 
\chi = \frac{\phi}{\sqrt{N} \;  M_{Pl}}  \; ,
\ee 
$w(\chi) \sim \mathcal{O}(1)~,~N\sim 50$ is the number of efolds
since the cosmologically relevant modes exited the horizon till the 
end of inflation and $ M $ is the energy scale of inflation

\medskip

The dynamics of the rescaled field $ \chi $ exhibits the slow
time evolution in terms of the \emph{stretched}
dimensionless time variable, 
\be \label{tau} 
\tau =  \frac{t \; M^2}{M_{Pl} \; \sqrt{N}}  \; .
\ee 
The rescaled variables $ \chi $ and $ \tau $ change slowly with time. 
A large change in the field amplitude $\phi$ results in a small change 
in the $ \chi $ amplitude, a change in $\phi \sim  M_{Pl}$ results in a 
$\chi$ change $\sim 1/\sqrt{N}$. The form of the potential, eq.(\ref{V}) 
and the rescaled dimensionless inflaton field   eq.(\ref{chifla}) and time 
variable $ \tau $ make manifest
the slow-roll expansion as a consistent systematic expansion in powers of 
$1/N$ \cite{1sN}.  

\medskip

The inflaton mass around the minimum is given by a see-saw formula
$$
m = \frac{M^2}{M_{Pl}}  \sim 2.45 \times 10^{13} \, \textrm{GeV} \; .
$$
The Hubble parameter when the cosmologically relevant modes exit the horizon
is given by
$$
H  = \sqrt{N} \; m \, {\cal H} \sim 1.0 \times 10^{14}\,\textrm{GeV}
= 4.1 \; m \; ,
$$
where we used that $  {\cal H} \sim 1 $. As a result, $ m\ll M $ and 
$ H \ll M_{Pl} $. A Ginsburg-Landau realization of the inflationary potential 
that fits the amplitude of the CMB anisotropy 
remarkably well, reveals that the
Hubble parameter, the inflaton mass and non-linear couplings are
see-saw-like, namely  powers of the ratio $M/M_{Pl}$ multiplied by
further powers of $ 1/N $. Therefore, the smallness of the couplings is not 
a result of fine tuning but a {\bf natural} consequence of the form of
the potential and the validity of the effective field theory
description and slow roll. The quantum expansion in loops is
therefore a double expansion on $ \left(H/M_{Pl}\right)^2 $ and $
1/N $. Notice that graviton corrections are  also at least of
order $ \left(H/M_{Pl}\right)^2 $ because the amplitude of tensor
modes is of order $H/M_{Pl}$ . We showed that the form of the potential which 
fits the WMAP data and is consistent with slow roll 
eqs.(\ref{V})-(\ref{chifla}) implies the small values for the inflaton 
self-couplings \cite{1sN}. 

\medskip

The equations of motion in terms of the dimensionless  rescaled field 
$ \chi $ and the slow time variable $ \tau $ take the form,
\bea \label{evol} 
&&  {\cal H}^2(\tau) = \frac13\left[\frac1{2\;N} 
\left(\frac{d\chi}{d \tau}\right)^2 + w(\chi) \right] \quad , \cr \cr
&& \frac1{N} \;  \frac{d^2
\chi}{d \tau^2} + 3 \;  {\cal H} \; \frac{d\chi}{d \tau} + w'(\chi) = 0 \quad .
\eea 
The slow-roll approximation follows by neglecting the
$\frac1{N}$ terms in eqs.(\ref{evol}). Both
$w(\chi)$ and $h(\tau)$ are of order $N^0$ for large $N$. Both
equations make manifest the slow roll expansion as an expansion in
$1/N$.

The number of e-folds $ N[\chi] $ since the field $ \chi $ exits the horizon 
till the end of inflation (where $ \chi $ takes the value $ \chi_{end} $) 
can be computed in close form from eqs. (\ref{evol}) in the slow-roll 
approximation (neglecting $ 1/N $ corrections)
\be \label{Nchi}
\frac{N[\chi]}{N} = -\int_{\chi}^{\chi_{end}}  \;
\frac{w(\chi)}{w'(\chi)} \; d\chi \;  \leqslant 1 \; .
\ee

\medskip

The amplitude of adiabatic scalar perturbations is expressed as
\cite{libros,WMAP,1sN,barrow,hu}
\be \label{ampliI}
|{\Delta}_{k\;ad}^{(S)}|^2  = \frac{N^2}{12 \, \pi^2} \;
\left(\frac{M}{M_{Pl}}\right)^4 \; \frac{w^3(\chi)}{w'^2(\chi)} \; .
\ee
The spectral index $ n_s $,  its running and the ratio of tensor to scalar 
fluctuations are expressed as
\bea \label{indi}
&&n_s - 1 = -\frac3{N} \; \left[\frac{w'(\chi)}{w(\chi)} \right]^2
+  \frac2{N}  \; \frac{w''(\chi)}{w(\chi)} \quad , \cr \cr 
&&\frac{d n_s}{d \ln k}= - \frac2{N^2} \; \frac{w'(\chi) \;
w'''(\chi)}{w^2(\chi)} - \frac6{N^2} \; \frac{[w'(\chi)]^4}{w^4(\chi)}
+ \frac8{N^2} \; \frac{[w'(\chi)]^2 \; w''(\chi)}{w^3(\chi)}\quad , \cr \cr 
&&r = \frac8{N} \; \left[\frac{w'(\chi)}{w(\chi)} \right]^2 \quad .
\eea
In eqs.(\ref{Nchi})-(\ref{indi}) the field $ \chi $ is computed at horizon 
exiting. We choose $ N[\chi] = N = 50 $.

Since, $ w(\chi) $ and  $ w'(\chi) $ are of order one, we
find from eq.(\ref{ampliI})
\be\label{Mwmap}
\left(\frac{M}{M_{Pl}}\right)^2 \sim \frac{2
\, \sqrt{3} \, \pi}{N} \; |{\Delta}_{k\;ad}^{(S)}| \simeq  1.02
\times 10^{-5} \; .
\ee
where we used $ N \simeq 50 $ and the WMAP
value for $ |{\Delta}_{k\;ad}^{(S)}| = (4.67 \pm 0.27)\times
10^{-5} $ \cite{WMAP}. This fixes the scale of inflation to be
$$
M \simeq 3.19 \times 10^{-3} \; M_{PL} \simeq 0.77
\times 10^{16}\,\textrm{GeV} \; .
$$
This value pinpoints the scale of
the potential during inflation to be at the GUT scale suggesting a
deep connection between inflation and the physics at the GUT
scale in cosmological space-time.

\medskip

We see that $|n_s -1|$ as well as  the ratio $ r $ turn out to be of order 
$ 1/N_{efolds} $. This nearly scale invariance is a natural property of
inflation which is described by a quasi-de Sitter space-time geometry.
This can be understood intuitively as follows: 
the geometry of the universe is scale invariant during de Sitter stage 
since the metric takes in conformal time the form 
$$
ds^2 = \frac1{(H \; \eta)^2}\left[ (d \eta)^2 - (d \vec x)^2 \right] \; .
$$
Therefore, the primordial power generated is scale invariant except
for the fact that inflation is not eternal and lasts for $N_{efolds}$.
Hence, the  primordial spectrum is scale invariant up to $ 1/N_{efolds} $ 
corrections. The values $ n_s = 1, \; r = 0 $ and $ d n_s/d \ln k
= 0 $ correspond to a critical point as discussed in ref.\cite{1sN}.
This a gaussian fixed point around which the inflation model hovers
in the renormalization group sense with an almost scale invariant spectrum 
of scalar fluctuations during the slow roll stage.

\medskip

The WMAP results favoured single inflaton models and among them new and hybrid
inflation emerge to be preferable than chaotic inflation \cite{nos}.

We analyze in the subsequent sections new inflation and hybrid
inflation in its simple physical realizations within the 
Ginzburg-Landau approach (the trinomial potential)\cite{nos}.

\section{Spectral index $n_s$, ratio $r$ and running index 
$ \frac{d n_s}{d \ln k} $ for New Inflation with the Trinomial Potential}

We consider here the trinomial potential investigated in ref.\cite{nos}
\be\label{VN}
V(\phi)= V_0 - \frac{m^2}{2} \; \phi^2 + \frac{ m \; g }{3} \; \phi^3 + 
\frac{\lambda}{4}\; \phi^4 \; . 
\ee
where $ m^2 > 0 $ and $ g $ and $ \lambda $ are dimensionless couplings.

The corresponding dimensionless potential $ w(\chi) $ takes the form
\be\label{trino}
w(\chi) = -\frac12 \; \chi^2 + \frac{h}3 \; \sqrt{\frac{y}2} \; \chi^3 +
\frac{y}{32} \; \chi^4 + \frac2{y} \; F(h)  \; ,
\ee
where the quartic coupling $ y $ is dimensionless as well as 
the asymmetry parameter $ h $. The couplings in eq.(\ref{VN}) and 
eq.(\ref{trino}) are related by,
\be 
g = h \; \sqrt{\frac{y}{2 \; N}} \; \left( \frac{M}{M_{Pl}}\right)^2  
\qquad ,  \qquad 
\lambda  = \frac{y}{8 \; N} \; \left( \frac{M}{M_{Pl}}\right)^4 
\label{aco} \; . 
\ee 
and the constant $ F(h) $ is related to $ V_0 $ by
$$
\frac2{y} \; F(h) = \frac{V_0}{N \; M^4 } \; .
$$
The constant $ F(h) $ ensures that 
$ w(\chi_+) =  w'(\chi_+) = 0 $ 
at the absolute minimum $ \chi = \chi_+ =\sqrt{\frac8{y}} (\Delta + |h|) $ 
of the potential $ w(\chi) $. Thus, inflation does not run eternally.
$ F(h) $ is given by
$$
F(h) \equiv \frac83 \, h^4 + 4 \, h^2 + 1 + \frac83 \, |h| \, \Delta^3 
\quad , \quad \Delta \equiv \sqrt{h^2 + 1} \; .
$$
The parameter $ h $ reflects how asymmetric is the potential.
Notice that  $ w(\chi) $ is invariant under the changes
$ \chi \to - \chi , \;  h \to - h $. Hence, we can restrict
ourselves to a given sign for $ h $. Without loss of 
generality, we choose $ h < 0 $ and shall
work with positive fields $ \chi $.

\begin{figure}[p]
\begin{turn}{-90}
\centering
\psfrag{ufa}{$n_s  ~ vs. ~ \log y$}
\includegraphics[width=10cm,height=14cm]{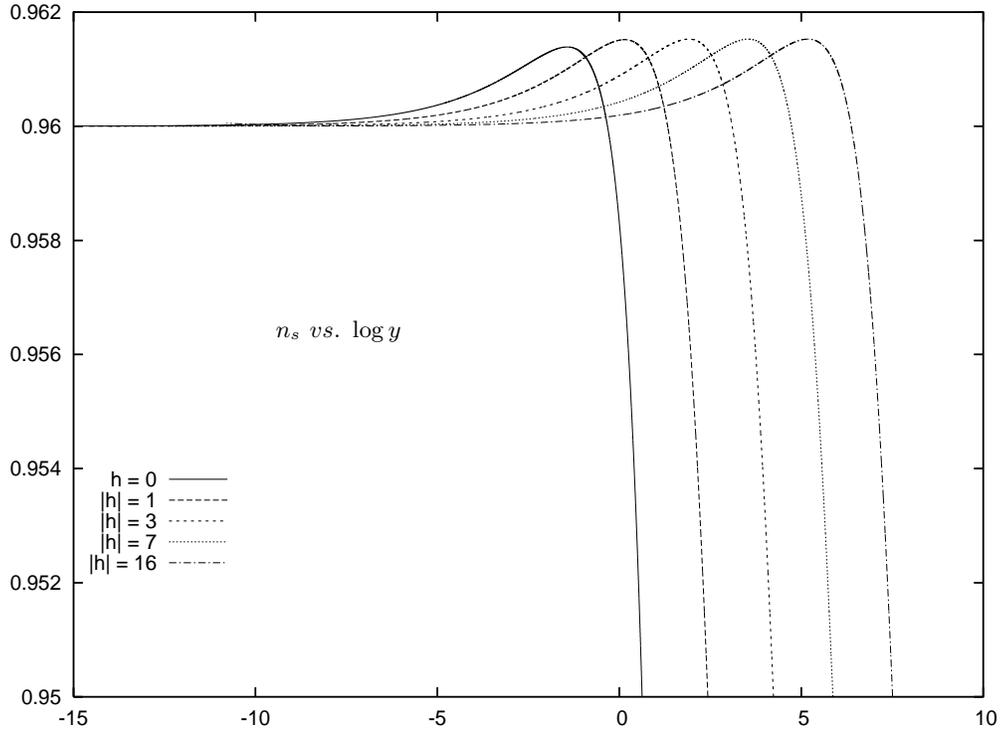}
\end{turn}
\caption{New Inflation. $ n_s $ as a function of $ \log y $ for the asymmetry 
of the potential $ |h| = 0, 1, 3, 7 $ and $ 16 $, $ y $ being the 
dimensionless quartic coupling. The $ y \to 0 $ limiting value $ n_s = 
1 - \frac{2}{N} = 0.96 $ is $h$-independent and corresponds to the monomial 
potential $ \frac12 \; m^2 \; \phi^2 $.}
\label{ns}
\end{figure}
\begin{figure}[p]
\begin{turn}{-90}
\centering
\includegraphics[width=10cm,height=14cm]{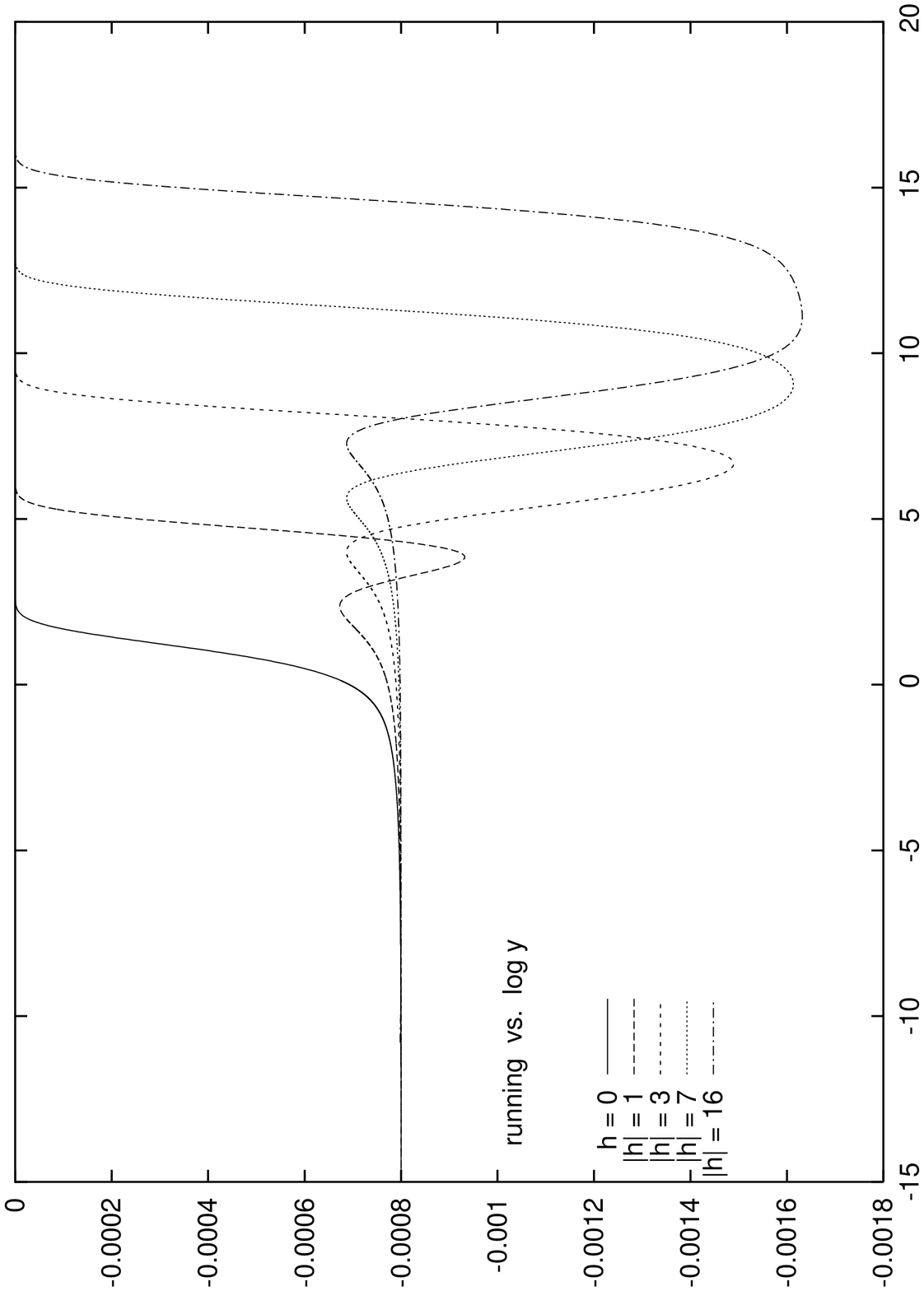}
\end{turn}
\caption{New Inflation. The running $ d n_s/d \ln k $ as a 
function of $ \log y $ for the 
asymmetry of the potential $ |h| = 0, 1, 3, 7 $ and $ 16 $,
$ y $ being the dimensionless quartic coupling.The $ y \to 0 $ limiting value 
$ -\frac2{N^2}= -0.0008 $ is $h$-independent and corresponds to the monomial 
potential $ \frac12 \; m^2 \; \phi^2 $.}
\label{frun}
\end{figure}

Notice that $ y \sim {\cal O}(1) \sim h $ guarantee that $ g  \sim 
{\cal O}(10^{-6}) $ and $ \lambda  \sim {\cal O}(10^{-12}) $ without
any fine tuning as stressed in ref. \cite{1sN}. 

New inflation is obtained by choosing the initial field $ \chi $ in 
the interval $ (0,\chi_+) $. The inflaton  $ \chi $ slowly rolls down the slope
of the potential from its initial value till the absolute minimum of the 
potential $ \chi_+ $.

Computing the number of efolds from eq.(\ref{Nchi}), we find the field 
$ \chi $ at horizon crossing related to the parameters $ y $ and $ h $.
It is convenient to define the field variable $z$:
$$
z \equiv \frac{y}8 \; \chi^2 \; .
$$
We obtain by inserting eq.(\ref{trino}) for $ w(\chi) $ into  eq.(\ref{Nchi})
and setting $ N[\chi] = N $,
\bea\label{ntrino}
&& y  = z - 2 \; h^2 -1 - 2  \; |h|  \; \Delta + \frac43 \; 
|h|  \; \left( |h| + \Delta - \sqrt{z} \right) + \cr \cr
&&+\frac{16}{3} \; |h| \;  (\Delta + |h| ) \; \Delta^2  \; 
\log\left[\frac12 \left(1 +  \frac{\sqrt{z} -  |h|}{\Delta}\right)\right] - 
2 \, F(h) \, \log\left[\sqrt{z} \; (\Delta - |h|)\right] \; .
\eea
$ z $ turns to be a monotonically decreasing function of $ y $:
$ z $ decreases from $ z = z _+ = (\Delta + |h|)^2$ till $ z = 0 $ when
$ y $ increases from $ y = 0 $ till $ y = \infty $.
When  $ \sqrt{z} \to 
\sqrt{z_+} , \; y $ vanishes quadratically,
$$
y \buildrel{z \to z_+}\over= 2 \; 
\left(\sqrt{z} - \sqrt{z_+}\right)^2 + {\cal O} 
\left(\left[\sqrt{z} - \sqrt{z_+}\right]^3\right) \; .
$$
We obtain in analogous way from eqs.(\ref{ampliI}) and (\ref{indi}) the 
spectral index, its running, the ratio $r$ and the amplitude of adiabatic 
perturbations,
\bea
&& n_s=1 - 6 \,  \frac{y}{N} \, \frac{z \; (z + 2  \, h  \, 
\sqrt{z} -1)^2}{\left[F(h)  -2 \, z+ \frac83  \, h  \,  z^{3/2} +
z^2\right]^2} +  \frac{y}{N} \, \frac{ 3 \, z+ 4 \, h  \, \sqrt{z} -1}{
F(h)- 2 \, z+ \frac83  \, h  \,  z^{3/2} + z^2} \quad , \\ \cr\cr
\label{nstrino}
&&\frac{d n_s}{d \ln k}= - \frac2{N^2} \; \sqrt{z} \; y^2 \; 
\frac{(z + 2  \, h  \, \sqrt{z} -1)(h + \frac32 \; \sqrt{z})}{
\left[F(h)  -2 \, z+ \frac83  \, h  \,  z^{3/2} +z^2 \right]^2} \cr\cr
&&- \frac{24}{N^2} \;  y^2 \; z^2 \; \frac{(z + 2  \, h  \, \sqrt{z} -1)^4}{
\left[F(h)  -2 \, z+ \frac83  \, h  \,  z^{3/2} +z^2 \right]^4} \cr\cr
&& + \frac8{N^2} \;  y^2 \; z \; \frac{ (3 \, z+ 4 \, h  \, \sqrt{z} -1)
(z + 2  \, h  \, \sqrt{z} -1)^2}{\left[F(h)  -2 \, z+ \frac83  \, h  \,  
z^{3/2} +z^2 \right]^3} \quad , \label{run} \\ \cr \cr
&& r = 16 \,   \frac{y}{N} \, \frac{z \; (z + 2  \, h  \, \sqrt{z} -1
)^2}{\left[F(h)  -2 \, z+ \frac83  \, h  \,  z^{3/2} +z^2
\right]^2}  \quad  , \label{rtrino}\\ \cr\cr
&&|{\Delta}_{k\;ad}^{(S)}|^2  = \frac{N^2}{12 \, \pi^2} \; 
\left(\frac{M}{M_{Pl}}\right)^4 \;
\frac{\left[F(h)- 2 \, z+ \frac83  \, h  \,  z^{3/2} + 
z^2\right]^3}{y^2 \; z \; (z + 2  \, h  \, \sqrt{z} -1)^2} \; .\label{dtrino}
\eea

\section{New Inflation with the Trinomial potential confronted to the 
three years WMAP data}

\begin{figure}[p]
\begin{turn}{-90}
\centering
\includegraphics[width=9cm,height=14cm]{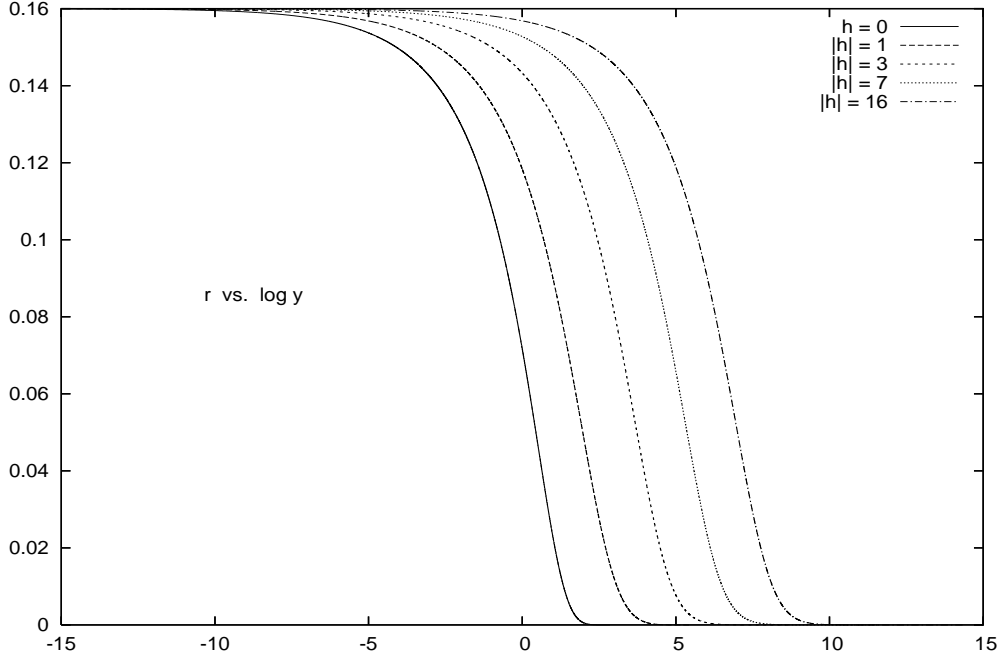}
\end{turn}
\caption{New Inflation. $ r $ as a function of $ \log y $ for the asymmetry 
of the potential $ |h| = 0, 1, 3, 7 $ and $ 16, \; y $ being the dimensionless 
quartic coupling. The absolute maximun value $ r = \frac8{N} = 0.16 $ is 
reached for
$ y = 0 $ and all $h$ and corresponds to the monomial potential 
eq.(\ref{maslim}).}
\label{r}
\end{figure}

\begin{figure}[p]
\begin{turn}{-90}
\centering
\psfrag{"nsr1.dat"}{$h = 0$} 
\psfrag{"nsr2.dat"}{$|h| = 0.15$} 
\psfrag{"nsr3.dat"}{$|h| = 0.4$} 
\psfrag{"nsr4.dat"}{$|h| = 0.7$} 
\psfrag{"nsr5.dat"}{$|h| = 20$} 
\psfrag{r  vs.  ns}{$r$ vs. $n_s$}
\includegraphics[width=10cm,height=14cm]{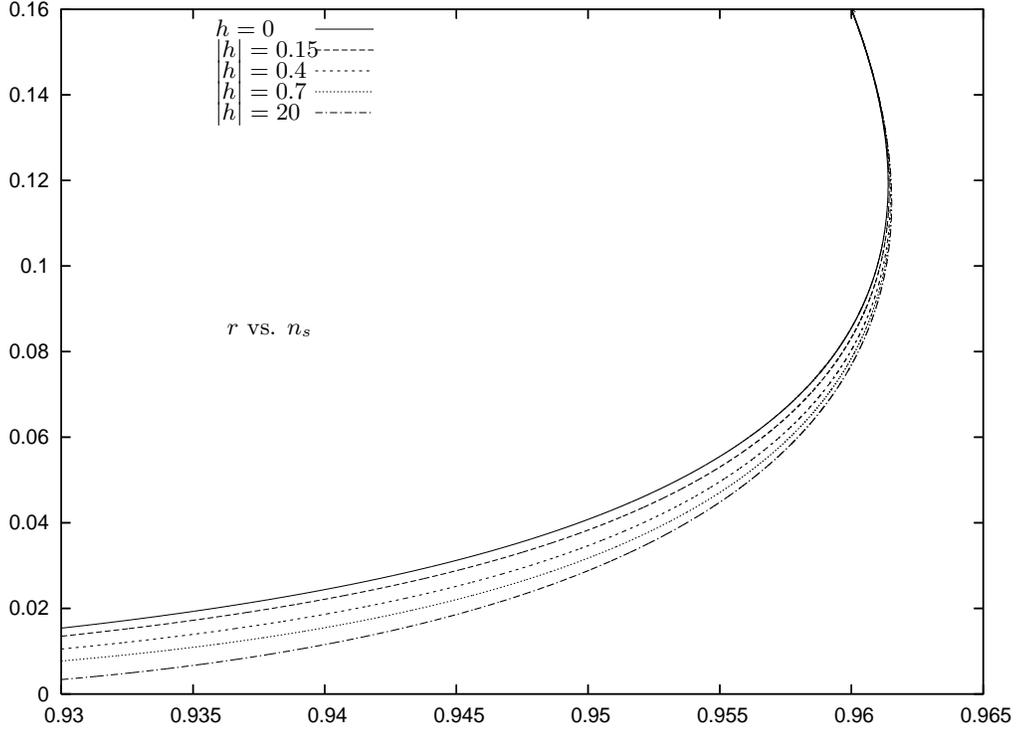}
\end{turn}
\caption{New Inflation. $ r $ as a function of $ n_s $ for the asymmetry 
of the potential $ |h| = 0, 0.15, 0.4, 0.7 $ and $ 20 $. For a given  $ n_s $,
$ r $ monotonically and slowly decreases with increasing $ |h| $.
$ r = r(n_s) $ is not too sensitive to $ h $. The maximun value of $ n_s $
is $ n_s^{maximum} = 0.961528\ldots $ and the corresponding $ r $ is
$ r_{max} =  0.114769\ldots $. The maximun value of  $ r $ is 
$ r_{abs~max} = 0.16 $ and corresponds to the quadratic potential setting 
$ y = 0 $ in eq.(\ref{trino}). For $ n_s = 0.95 $ 
(the three years WMAP value), we find $ 0.03 < r < 0.04 $.}
\label{nsr}
\end{figure}

\begin{figure}[p]
\begin{turn}{-90}
\centering
\psfrag{"nsrun1.dat"}{$h = 0$} 
\psfrag{"nsrun2.dat"}{$|h| = 0.15$} 
\psfrag{"nsrun3.dat"}{$|h| = 0.4$} 
\psfrag{"nsrun4.dat"}{$|h| = 0.7$} 
\psfrag{"nsrun5.dat"}{$|h| = 20$} 
\includegraphics[width=10cm,height=14cm]{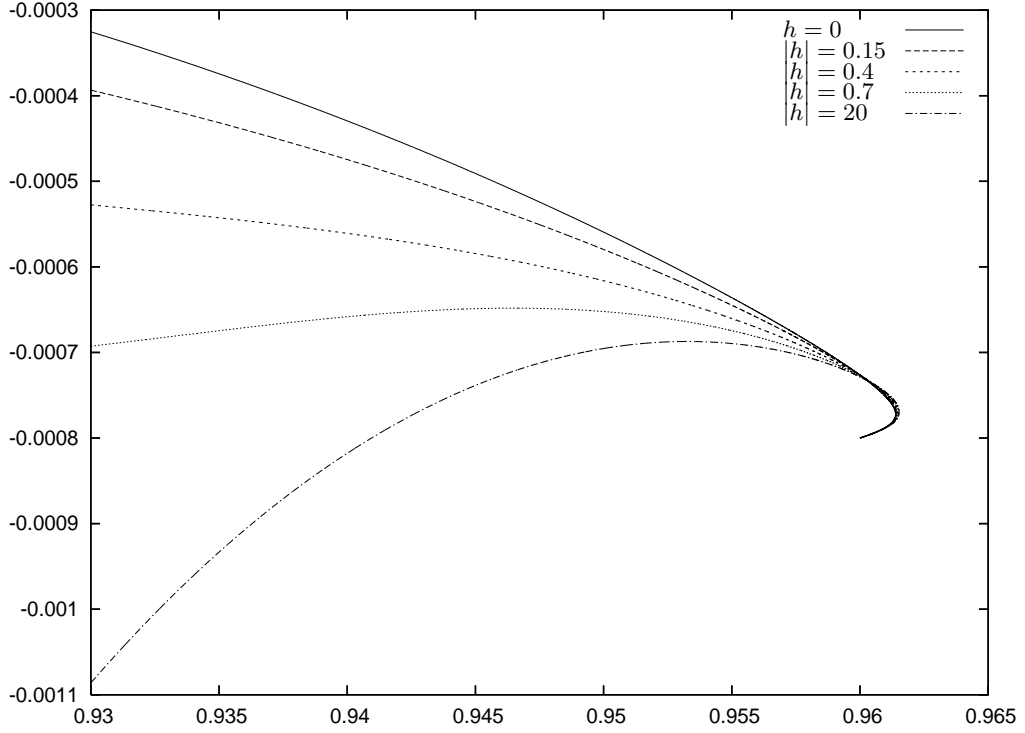}
\end{turn}
\caption{New Inflation. The running $ d n_s/d \ln k $ as a function of $ n_s $ 
for the asymmetry of the potential $ |h| = 0, 0.15, 0.4, 0.7 $ and $ 20 $.
The running turns out to be always {\bf negative} in new inflation.
For $ n_s < 0.96 $, the running $ d n_s/d \ln k $ decreases with increasing
$ |h| $. The opposite happens for  $ n_s > 0.96 $. In the last case the 
dependence on $ h $ is weak. We find $ d n_s/d \ln k = -0.00077\ldots $ at the 
branch point $ n_s = 0.961\ldots $ for all values of $|h|$. 
The point $ n_s = 1 - \frac{2}{N} = 0.96, \;  \frac{d n_s}{d \ln k} 
= -\frac2{N^2}= -0.0008 $ is reached for 
all values of $ h $ and corresponds to the monomial potential 
eq.(\ref{maslim}). For $ n_s = 0.95 $ (the three years WMAP value), 
we find $  -0.00070 < d n_s/d \ln k < -0.00055 $.}
\label{nsrun}
\end{figure}

We plot $ n_s $, its running and $ r $ in figs. \ref{ns}, \ref{frun} 
and \ref{r} as functions of $ \log y $ for various values of the asymmetry 
of the potential $ h , \; y $ being the dimensionless quartic coupling.
Figs. \ref{nsr} and \ref{nsrun} depict $ r $ and the running 
$ d n_s/d \ln k $ as functions of $ n_s $ for various values of the 
asymmetry $ h $. 

\begin{figure}[p]
\begin{turn}{-90}
\centering
\psfrag{maxima of n_s vs. |h|}{maxima of $n_s$ vs. $|h|$}
\includegraphics[width=9cm,height=14cm]{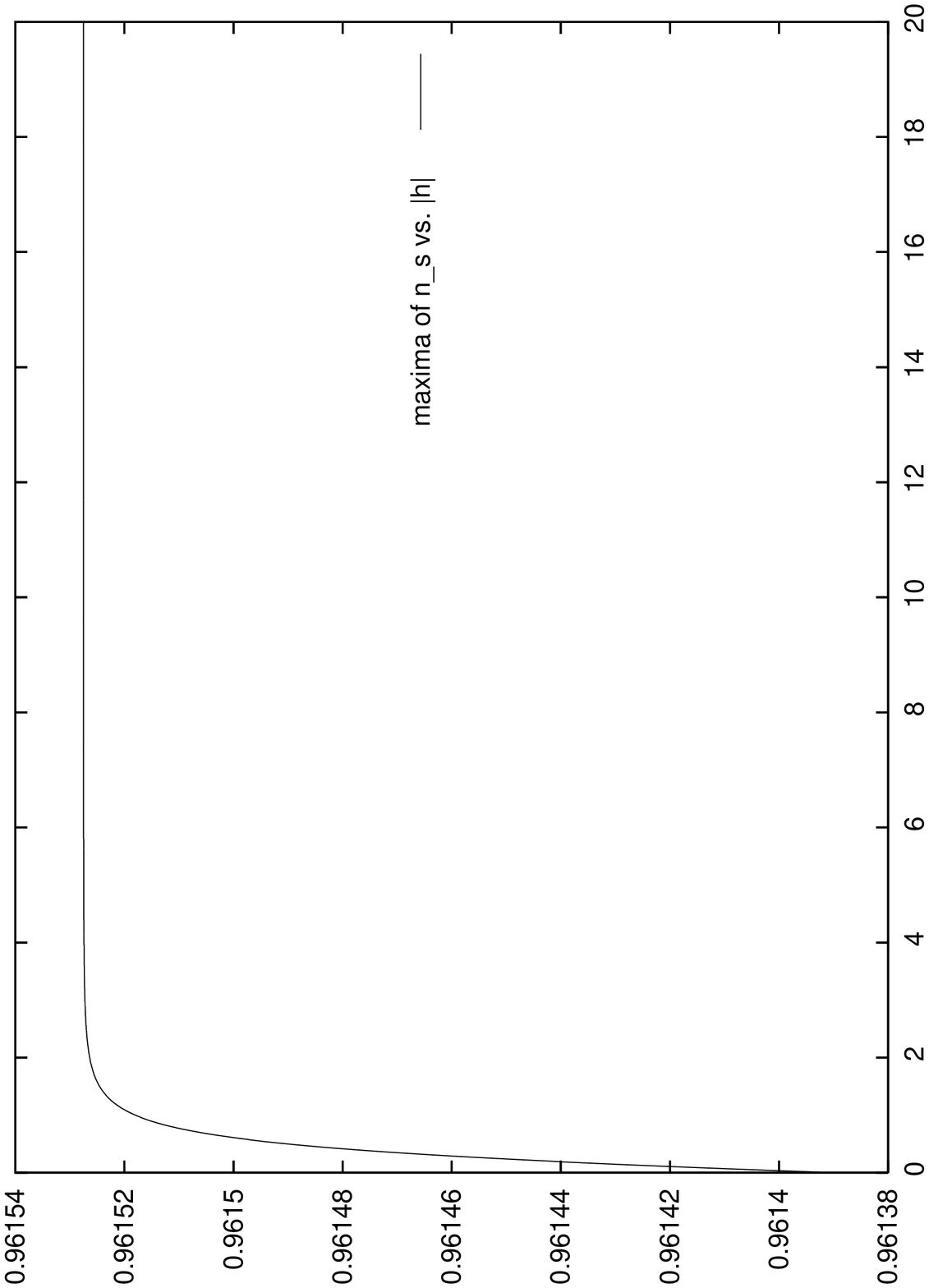}
\end{turn}
\caption{New Inflation. Maxima of $ n_s $ plotted vs. the asymmetry 
of the potential $ |h| $. The limiting value for large $ |h| $ is 
$ n_s^{maximum} = 0.961528\ldots $}
\label{nsh}
\end{figure}

We see that generically $ n_s < 1 $ and  $ d n_s/d \ln k < 0 $ for new 
inflation for all values of the couplings. 

In new inflation we have the absolute upper bound 
\be \label{cotar}
{\rm new~inflation:} \qquad r \leq  r_{abs~max} = \frac{8}{N} = 0.16 \quad  ,
\ee
which is attained by the quadratic monomial potential obtained from 
eq.(\ref{trino}) at $ y = 0 $. On the contrary, in chaotic inflation $ r $ 
is bounded as
$$
{\rm chaotic~inflation:} \quad 0.16 = \frac8{N} < r < \frac{16}{N}= 0.32 \quad .
$$
This bound holds for all values the asymmetry parameter $ h $.
The lower and upper bounds for $ r $ are saturated by the quadratic and quartic
monomials, respectively.

We see from fig. \ref{ns} that $ n_s $ exhibits a single maximun 
$ n_{s~maximum}(h) $ as a function of the quartic coupling $ y $ for fixed 
asymmetry $ h $. In fig. \ref{nsh} we plot $ n_{s~maximum}(h) $ as a function
of $ h $.  $ n_{s~maximum}(h) $ monotonically increases with $ |h| $ and
rapidly reaches its limiting value $ n_{s~maximum} = 0.961528 \ldots $.
The corresponding value for $ r $ is $ r =  0.114769\ldots $.
Values $ n_s >  n_{s~maximum}  = 0.961528 \ldots $ cannot be described by 
new inflation with the trinomial potential eqs.(\ref{VN})-(\ref{trino}).

We see from fig.  \ref{nsr} and  \ref{nsrun} that both $ r $ and the running 
$ d n_s/d \ln k $ are {\bf two}-valued functions of $ n_s $ in the interval 
$ 0.96 < n_s < 0.961528 \ldots $. That is, for each  $ n_s $ in this range 
there are  {\bf two} possible values for $ r $ and for the running 
$ d n_s/d \ln k $.
Therefore, we can cover the whole range of values $ 0.96 < n_s < 0.961528 
\ldots $ choosing the lower branch for $ r $. We find for this branch 
$$ 
r <  r_{max} \equiv 0.114769\ldots \; . 
$$
This maximun value $ r_{max} $ is well below the absolute
maximun in new inflation $ r_{abs~max} = 0.16 $ [eq.(\ref{cotar})] which
belongs to the second branch.

\medskip

The plots of the ratio $ r $ and the running $ d n_s/d \ln k $ as a 
function of $ n_s $ show that these quantities are not very sensitive
to the asymmetry $ h $ for a given value of $ n_s $.

\bigskip

The three years WMAP \cite{WMAP3} data as well as ref. \cite{2dF}
yield for $ n_s $ the value (see also refs. \cite{SDSS} and \cite{Teg})
\be \label{nswmap}
 n_s = 0.95 \pm 0.02 \; .
\ee
For $ n_s = 0.95 $ and any value of the asymmetry $ h $ [see fig. \ref{nsr}],
new inflation with the trinomial potential eqs.(\ref{VN})-(\ref{trino})
yields 
$$
{\rm new~inflation:} \quad 0.03 < r < 0.04 \quad {\rm and}
\quad -0.00070 < d n_s/d \ln k < -0.00055 \; .
$$
New inflation with the trinomial potential always yield $ n_s $ below
the maximun value $ n_{s~maximum} = 0.961528 \ldots $. 
For $ n_s = 0.961528 \ldots $ we have in this model $ r =  0.114769\ldots $.
These values of the ratio $ r $ fullfil the three years WMAP bound
including SDSS galaxy survey \cite{WMAP3}
\be \label{rwmap}
r < 0.28 \; (95\% ~CL) \; .
\ee
Moreover, one can see from fig. 14 in ref. \cite{WMAP3} 
that $ r < 0.1 ~(68\% ~CL) $ from WMAP$+$SDSS. 

\medskip

Chaotic inflation  with the trinomial potential eq.(\ref{VN})-(\ref{trino})
yields larger values of $ r $ than new inflation
for a given value of $ n_s $ \cite{nos}. More precisely, 
we find $ r = 0.27 $ for $ n_s = 0.95 $ for the binomial potential in
chaotic inflation \cite{nos}
(the trinomial potential introduces very small changes).

Therefore, although the WMAP value for $ n_s $ [eq.(\ref{nswmap})]
is compatible both with chaotic and new inflation, the WMAP bounds on 
$ r $ {\bf clearly disfavour} chaotic inflation. New inflation easily fulfils 
the three years WMAP bounds on $ r $.

\medskip

The present data do not permit to find the precise values neither
of the ratio $ r $ nor of the running index $ d n_s/d \ln k $; 
only upper bounds are obtained \cite{WMAP,WMAP3}. We therefore think
that the value of $ n_s $ [eq.(\ref{nswmap})] obtained through a fit of the data
assuming $ r = 0 $ is more precise than the values of $ n_s $ obtained
through fits allowing both $ r $ and $ d n_s/d \ln k $ to vary.
Notice that $ n_s = 0.95 $ was independently found from the 2dF data
under similar assumptions \cite{2dF}.

Ref. \cite{WMAP3} reports fits  
yielding negative values for $ d n_s/d \ln k $ of the order $ \sim -0.05 $.
Notice that the order of magnitude of the running $ d n_s/d \ln k $ 
is just fixed by the fact that it is a second order quantity in slow-roll:
$ \sim \frac1{N^2} \sim 0.0004 $. Still, the negative sign of the running
reported by ref.\cite{WMAP3} agrees with the sign prediction of new inflation 
with the trinomial potential [see fig. \ref{frun} and \ref{nsrun}]. 

\medskip

In summary, new inflation with the trinomial potential 
eq.(\ref{VN})-(\ref{trino}) predicts $ 0.03 < r < 0.04 $ and $ -0.00070 < 
d n_s/d \ln k < -0.00055 $ for $ n_s = 0.95 $. For $ 0.93 <  n_s < 0.962 $
it predicts $  0.01 < r < 0.115 $ and $ -0.001 < d n_s/d \ln k < -0.0003 $
[see figs. \ref{nsr} and \ref{nsrun}].

\section{Limiting Cases of the Trinomial Potential in New Inflation}

Let us now consider the limiting cases: the shallow limit  ($ y \to 0 $),
 the steep limit $ y \to \infty $ and the extremely asymmetric limit $ |h| \to 
\infty $ of the trinomial potential for new inflation 
eqs.(\ref{VN})-(\ref{trino}).

\subsection{The shallow limit $ y \to 0 $ of the Trinomial Potential}

In the shallow  limit $ y \to 0 , z $ tends to $ z = z_+ = (\Delta + |h|)^2 $,
 which is the minimum of $ y $ in eq.(\ref{ntrino}).
We find from eqs.(\ref{ntrino})-(\ref{dtrino}), 
\bea\label{cotrino}
&& n_s \buildrel{y \to 0}\over= 1 - \frac{2}{N} \simeq 0.96  
\quad , \quad \frac{d n_s}{d \ln k} \buildrel{y \to 0}\over= 
-\frac2{N^2}\simeq -0.0008 \quad , \cr \cr
&& r \buildrel{y \to 0}\over= \frac{8}{N}  \simeq 0.16 
\quad , \quad 
|{\Delta}_{k\;ad}^{(S)}|^2  \buildrel{y \to 0}\over= \frac{N^2}{3 \, \pi^2} \; 
\left(\frac{M}{M_{Pl}}\right)^4 \; \Delta(\Delta+|h|) \; ,
\eea
which coincide with $ n_s , \; \frac{d n_s}{d \ln k}$ and $ r $ 
for the monomial quadratic potential. That is, the  $ y \to 0 $ limit
is $h$-independent except for $|{\Delta}_{k\;ad}^{(S)}|$.
For fixed $ h $ and $ y \to 0 $ the inflaton potential eq.(\ref{trino}) 
becomes purely quadratic:
\be \label{maslim}
w(\chi) \buildrel{y \to 0}\over= \Delta ( \Delta + |h| ) \; (\chi -\chi_+)^2 +
{\cal O}(\sqrt{y} )  \; ,
\ee
where $ \chi_+ \equiv \sqrt{\frac8{y}} \; ( \Delta + |h| ) $.
Notice that the amplitude of scalar adiabatic fluctuations eq.(\ref{cotrino})
turns out to be proportional to the square mass of the inflaton in this 
regime which we read from eq.(\ref{maslim}): $ 2 \; \Delta ( \Delta + |h| ) $. 
The shift of the inflaton field by $ \chi_+ $ has no observable consequences.

The numerical values in eq.(\ref{cotrino}) are in agreement with
figs. \ref{ns}-\ref{r} in the  $ y \to 0 $ limit. For $ h = 0 $ we recover the
results of the monomial potential. 

\subsection{The  steep limit $ y \to \infty $ of the Trinomial Potential}

In the  steep limit $ y \to \infty, \;  z $ tends to zero for new inflation.
We find from eq.(\ref{ntrino})
\be\label{trikG}
y  \buildrel{z \to 0}\over=- F(h) \; \log z
-q(h) -1 + {\cal O}(\sqrt{z}) \quad  ,
\ee
where
$$
q(h) \equiv 2 \,  F(h) \log[\Delta-|h|] -
\frac23 \; \left( h^2 + |h| \; \Delta \right) 
\left\{ 8 \, \Delta^2 \, 
\log\left[\frac12 \left(1 - \frac{|h|}{\Delta}\right)\right] - 1 \right\}
\; ,
$$
$ q(h) $ is a monotonically increasing function of the asymmetry 
$ |h| : \; 0 \leq q(h) < \infty $ for $ 0 < |h| < \infty $.

Then, eqs.(\ref{nstrino})-(\ref{rtrino}) yield,
\bea\label{nsrtrikG}
&&n_s \buildrel{y \gg 1}\over=1 - \frac{y}{N \; F(h)}
\quad , \quad r \buildrel{y \gg 1}\over= 
\frac{16 \; y}{N \; F^2(h)} \, e^{- \frac{y+1+q(h)}{F(h)}}\quad ,
\cr \cr
&& \frac{d n_s}{d \ln k}\buildrel{y \gg 1}\over= -\frac{2 \; y^2 \; |h|}{N^2 \;
 F^2(h)} \, e^{- \frac{y+1+q(h)}{2 \; F(h)}}\quad , \cr \cr
&& |{\Delta}_{k\;ad}^{(S)}|^2  \buildrel{y \gg 1}\over= 
\frac{N^2}{12 \, \pi^2} \; \left(\frac{M}{M_{Pl}}\right)^4 \; 
\frac{F^3(h)}{y^2} \; e^{\frac{y+1+q(h)}{F(h)}} \quad .
\eea
In the $ h \to 0 $ limit we recover from eqs.(\ref{trikG})-(\ref{nsrtrikG})
the results for new inflation with a purely quartic potential:
we have $ F(0) = 1 $ and $ q(0) = 0 $ and eq.(\ref{nsrtrikG}) becomes,
\bea\label{nsrtrikGh0}
&&n_s \buildrel{y \gg 1, h \to 0}\over=1 - \frac{y}{N}
\quad , \quad r \buildrel{y \gg 1, h \to 0}\over= 
\frac{16 \; y}{N} \, e^{-y-1}\quad ,
\cr \cr
&& \frac{d n_s}{d \ln k}\buildrel{y \gg 1, h \to 0}\over= 
-\frac{2 \; y^2 \; |h|}{N^2} \, e^{-y-1}\quad , \cr \cr
&& |{\Delta}_{k\;ad}^{(S)}|^2  \buildrel{y \gg 1, h \to 0}\over= 
\frac{N^2}{12 \, \pi^2} \; \left(\frac{M}{M_{Pl}}\right)^4 \; 
\frac{e^{y+1}}{y^2} \;  \quad .
\eea
The behaviour in eqs.(\ref{nsrtrikG}) is in agreement with 
figs. \ref{ns}-\ref{r} in the  $ y \to +\infty $ limit.

\subsection{The extremely asymmetric limit $ |h| \to \infty $ 
of the Trinomial Potential}

Eqs.(\ref{ntrino})-(\ref{dtrino}) have a finite limit for $ |h| \to \infty 
$ with $ y $ and $ z $ scaling as $ h^2 $. Define,
$$
Z \equiv \frac{z}{h^2} \quad , \quad Y \equiv \frac{y}{h^2} \; .
$$
Then, we find for $ |h| \to \infty $ 
from eqs.(\ref{ntrino})-(\ref{dtrino}) keeping $ Z $ and $ Y $ fixed,
\bea \label{hgra}
&& Y = Z - \frac43 \; \sqrt{Z} - 4 - \frac43 \; \log\frac{Z}4 +
\frac{16}{3 \; \sqrt{Z}}\; , \cr \cr
&& n_s = 1 - 6 \; \frac{Y}{N} \; \frac{Z^2 \; (\sqrt{Z} - 2)^2}{
[\frac{16}3 - \frac83 \; Z^{\frac32} + Z^2 ]^2 } + \frac{Y}{N} \;
\frac{ 3 \; Z - 4 \; \sqrt{Z} }{\frac{16}{3}  - \frac83 \; Z^{\frac32} + Z^2 }
\; , \\ \cr
&&\frac{d n_s}{d \ln k}= - \frac2{N^2} \;  Y^2 \; Z \; 
\frac{(\sqrt{Z} -2)(\frac32 \; \sqrt{Z}-1)}{
\left[\frac{16}3  - \frac83 \,  Z^{3/2} + Z^2 \right]^2} \cr\cr
&&- \frac{24}{N^2} \;  Y^2 \; Z^4 \; \frac{(\sqrt{Z} -2)^4}{
\left[\frac{16}3  - \frac83 \,  Z^{3/2} + Z^2 \right]^4} 
+ \frac8{N^2} \;  Y^2 \; Z^{\frac52} \; 
\frac{(3 \,\sqrt{Z} - 4)(\sqrt{Z} -2)^2}{
\left[\frac{16}3  - \frac83 \,  Z^{3/2} + Z^2 \right]^3}
\quad , \label{runh} \\ \cr \cr
&& r = 16 \; \frac{Y}{N} \; \frac{Z^2 \; (\sqrt{Z} - 2)^2}{
[\frac{16}{3} - \frac83 \; Z^{\frac32} + Z^2 ]^2 }  \; ,  \cr \cr
&& |{\Delta}_{k\;ad}^{(S)}|^2  = \frac{N^2 \; h^2}{12 \, \pi^2} \; 
\left(\frac{M}{M_{Pl}}\right)^4 \;
\frac{[\frac{16}{3}  - \frac83 \; Z^{\frac32} + Z^2]^2}{Y^2 \; Z^2  
\; (\sqrt{Z} - 2)^2} \; . \label{amplihg}
\eea
We have $ 0 \leq Z \leq 4 $ for $ +\infty \geq Y \geq 0 $. 
In the  $ |h| \to \infty $ limit the inflaton potential takes the form
$$
W(\chi) \equiv \lim_{|h| \to \infty} \frac{w(\chi)}{h^2} =
\frac{32}{3 \; Y} - \frac13 \; \sqrt{\frac{Y}{2}} \; \chi^3 
+ \frac{Y}{32} \; \chi^4 \; .
$$
This is a broken symmetric potential without quadratic term.
Notice that the cubic coupling has dimension of a mass in eq.(\ref{VN})
and hence this is {\bf not} a massless potential contrary to the quartic 
monomial $ \chi^4 $. In addition, eq.(\ref{amplihg}) shows that for large 
$ |h| $ one must keep the product $ |h| \; M^2 $ fixed
since it is determined by the amplitude of the adiabatic perturbations.
We see from eq.(\ref{amplihg}) that $ {\tilde M} \equiv \sqrt{|h|} \; M $ becomes 
the energy scale of inflation in the $ |h| \to \infty $ limit.
$ {\tilde M} \sim 10^{16}$GeV according to the observed value of 
$ |{\Delta}_{k\;ad}^{(S)}|/N $ displayed in eq.(\ref{Mwmap}), while
$ M $ and $ m $,
$$
M = \frac{\tilde M}{\sqrt{|h|}} \buildrel{ |h| \to \infty 
}\over= 0 \quad ,  \quad
m = \frac{M^2}{M_{Pl}} = \frac{ {\tilde M}^2}{|h| \; M_{Pl}} \buildrel{ |h| \to \infty 
}\over= 0 \quad .
$$
vanish as  $ |h| \to \infty $.

\medskip

The curves in figs. \ref{ns}-\ref{nsh}
for high values of $ |h| $ are well described by eq.(\ref{hgra}).

\section{Hybrid Inflation}

In the inflationary models of hybrid type, the inflaton is coupled to another 
scalar field $ \sigma_0 $ with mass term $ -\mu_0^2 < 0 $
through a potential of the type \cite{lin}
\bea\label{Vhib1}
&&V_{hyb}(\phi,\sigma_0) = \frac{m^2}{2} \; \phi^2 + \frac{g_0^2}{2} \; 
\phi^2 \; \sigma_0^2 + \frac{\mu_0^4}{16 \, \Lambda_0} 
\left(\sigma_0^2 -  \frac{4 \, \Lambda_0}{\mu_0^2} \right)^2= \cr \cr
&&  =\frac{m^2}{2} \; \phi^2 + \Lambda_0+\frac12 \; (g_0^2 \; 
\phi^2-\mu_0^2) \;  \sigma_0^2+\frac{\mu_0^4}{16 \; \Lambda_0} \; 
\sigma_0^4\; ,
\eea
where $ m^2 > 0 , \; \Lambda_0 > 0 $ plays the role of a cosmological constant 
and $ g_0^2 $  couples $ \sigma_0 $ with $ \phi $.

\medskip

The initial conditions are chosen such that $\sigma_0$ and $ \dot\sigma_0$ 
are very small (but not identically zero) and therefore one can consider 
initially,
\be\label{vsig0}
V_{hyb}(\phi,0) =\frac{m^2}{2} \; \phi^2 + \Lambda_0 \; .
\ee
One has then inflation driven by the cosmological constant $ \Lambda_0 $ 
plus the initial value of the inflaton $  \phi(0) $.
The inflaton field $ \phi(t) $ 
decreases with time while the scale factor $ a(t) $ grows exponentially 
with time. We see from eq.(\ref{Vhib1}) that
\be\label{masefe}
m_{\sigma}^2 = g_0^2 \; \phi^2 - \mu_0^2  \; ,
\ee
plays the role of a effective classical mass square for the field $ \sigma_0 $.
The initial value of $ m_{\sigma}^2 $ depends on the initial conditions
but is typically positive.
In any case, since  the inflaton field $ \phi $ decreases
with time, $ m_{\sigma}^2 $ will be necessarily negative at some 
moment during inflation.
At such moment, spinodal (tachyonic) unstabilities appear and the
field $ \sigma_0 $ starts to grow exponentially. 
Inflation stops when both fields $ \phi $ and $ \sigma_0 $ reach their
vaccum values. A matter dominated regime follows.

\bigskip

Normally, the field $ \sigma_0 $
is negligible when the relevant cosmological scales cross out the 
horizon. Hence, $ \sigma_0 $ does not affect the spectrum of density and
tensor fluctuations except through the number of efolds. 
Hence, hybrid inflation is a single-field inflationary model
as long as $ \phi $ solely contributes to the spectrum of density and tensor 
fluctuations. However, we find specific regions of parameters 
$ (g_0, \mu_0 , \Lambda_0 ) $ and initial 
conditions where both fields  $ \phi $ and $ \sigma_0 $ contribute to the 
cosmologically relevant fluctuations making hybrid inflation a two-field 
inflationary model. We do not consider such  regions of parameters here which 
are outside the scope of this paper.

\bigskip

In terms of the dimensionless fields and couplings, the potential  
$ V_{hyb}(\phi,\sigma_0) $  eq.(\ref{Vhib1}) reads
\bea
&& w(\chi,\sigma) = \frac12 \; \chi^2 
+ \frac{\mu^4}{8 \; \Lambda}\left(\sigma^2 - \frac{2 \, \Lambda}{\mu^2} 
\right)^2 + \frac12 \; g^2 \; \sigma^2 \; \chi^2 = \cr \cr
&&= \frac12 \; \chi^2 + \frac12 \; ( g^2 \; \chi^2-\mu^2 ) \; \sigma^2  +
\frac12 \;\Lambda + \frac{\mu^4}{8 \; \Lambda} \; \sigma^4 \; .
\eea
where 
$$ \sigma(\tau) \equiv \frac{\sigma_0(t)}{\sqrt{N} \; M_{Pl}} \quad , \quad
 g^2 \equiv g_0^2 \; \frac{N \; M_{Pl}^2}{m^2}\quad , \quad
\mu^2 \equiv \frac{\mu_0^2}{m^2} \quad  \mbox{and}  
\quad \Lambda \equiv \frac{2 \, \Lambda_0}{M^4 \; N} \; .
$$
The evolution equations for this potential in dimensionless variables 
take the form
\bea\label{ecmovh}
&& {\cal H}^2(\tau) = \frac1{6} \; \left[ \frac1{N} \; {\dot \chi}^2 + \chi^2 
+ \frac{\mu^4}{4 \; \Lambda}\left(\sigma^2 - \frac{2 \, \Lambda}{\mu^2} 
\right)^2 + g^2 \; \sigma^2 \; \chi^2
\right] \; , \cr \cr
&& 
\left[\frac1{N} \; \frac{d^2}{d \tau^2}  + 3 \,  {\cal H} \,\frac{d}{d \tau}
+ 1 + g^2 \; \sigma^2 \right] \chi = 0 \; , \label{fihib} \\
&& 
\left[\frac1{N} \; \frac{d^2}{d \tau^2}  + 3 \,  {\cal H} \,\frac{d}{d \tau}
 - \mu^2+ g^2 \; \chi^2 +\frac{\mu^4}{2 \; \Lambda} \; \sigma^2 \right]
\sigma(\tau) = 0  \; . \nonumber
\eea
Since the field $\sigma$ is chosen initially very small, it
can be neglected and we can approximate the evolution equations 
(\ref{fihib}) as
\be \label{ecmslo}
3 \,  {\cal H} \, {\dot \chi} + \chi = 0 \quad , \quad
 {\cal H}^2(\tau) = \frac1{6} \; \left[ \chi^2 + \Lambda  \right] \; .
\ee
The number of efolds from the time $\tau$ till the end of inflation
is then given by eq.(\ref{Nchi}),
\be\label{Nhib}
N(\tau) = N \; \int_{\tau}^{\tau_{end}}  {\cal H}(\tau) \; d\tau =  
-  \int_{\chi(\tau)}^{\chi_{end}} \frac{w(\chi)}{w'(\chi)} \; d\chi
= \frac{N}4 \left[ \chi^2(\tau)- \chi^2_{end} \right]
+ \frac{N}2 \; \Lambda \; \log\frac{\chi(\tau)}{\chi_{end}}  \; ,
\ee
$ \chi_{end} $ is the inflaton field at the end of inflation.

\medskip

We see that the inflaton field and its dynamics only appear in $ N(\tau) $ 
eq.(\ref{Nhib}) through the value of $ \chi_{end} $ where inflation stops. The
value of $ \chi_{end} $ follows by solving eqs.(\ref{ecmovh}) and 
depends on the initial conditions as well as on the parameters 
$ g, \; \mu $ and $ \Lambda $.

\medskip

The spectral indices are given by eqs.(\ref{indi}) and
the amplitude of adiabatic perturbations by eq.(\ref{ampliI}).
By using the potential eq.(\ref{vsig0}) in dimensionless variables we find,
\bea\label{indhi}
&& w(\chi,0) = \frac12 \, ( \chi^2 + \Lambda) \quad , \cr \cr
&& |{\Delta}_{k\;ad}^{(S)}|^2 = \frac{N^2}{96 \, \pi^2 } 
\left( \frac{M}{M_{Pl}} 
\right)^4 \; \frac{(\chi^2 + \Lambda)^3}{\chi^2} \quad ,  \quad 
r = \frac{32}{N} \; \frac{\chi^2}{(\chi^2 + \Lambda)^2}  \; ,  \\ \cr
&&n_s = 1 + \frac4{N} \; \frac{\Lambda - 2 \; \chi^2}{(\chi^2 + \Lambda)^2}
\quad ,  \quad \frac{d n_s}{d \ln k}= \frac{32}{N^2} \; 
\frac{\chi^2(2 \,\Lambda-\chi^2)}{(\chi^2 + \Lambda)^4} 
\label{nshib} \; , 
\eea
where $ \chi $ is the inflaton at the moment of the first horizon crossing.

\begin{figure}[p]
\begin{turn}{-90}
\centering
\includegraphics[width=10cm,height=14cm]{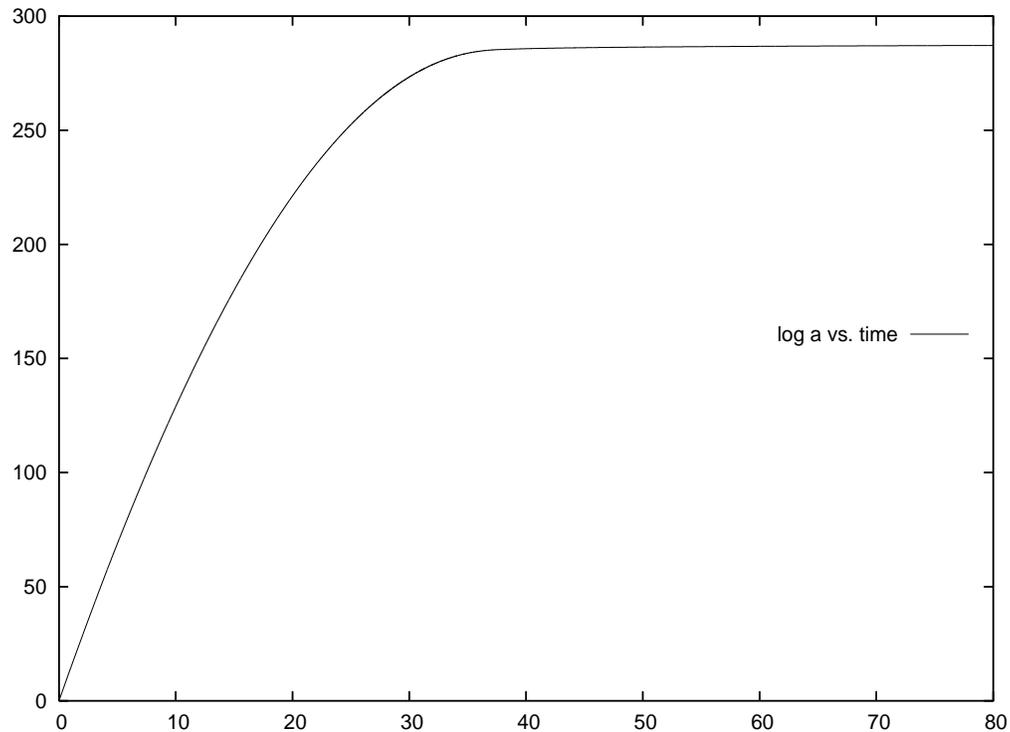}
\end{turn}
\caption{Hybrid inflation. The logarithm of the scale factor 
(the number of efolds) as a function of time. The chosen parameters in 
eq.(\ref{ecmovh}) are $ \Lambda = 4 \; N = 200 \;  , \; \mu^2 = 1.7 \; 
\Lambda, \; \phi(0) = 2.3 \; \sqrt{\Lambda} $. 
A stage of slow-roll quasi de Sitter
inflation takes place (till $ \tau \simeq 39 $ in this example) followed by
a matter dominated era.}
\label{a}
\end{figure}

\begin{figure}[p]
\begin{turn}{-90}
\centering
\includegraphics[width=10cm,height=14cm]{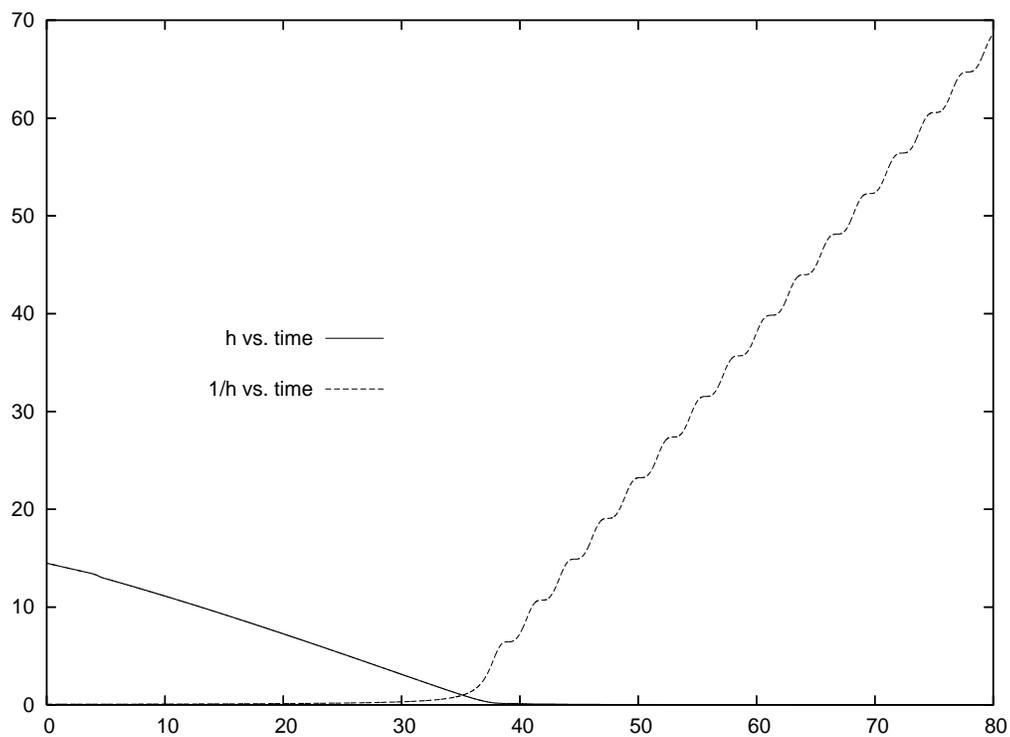}
\end{turn}
\caption{ Hybrid inflation. The Hubble parameter $ h $ and its inverse $ 1/h $ 
as a function of time. Same parameters as in fig. \ref{a}.  $ h $ slowly
decreases with time in the slow-roll quasi de Sitter stage 
(till $ \tau \simeq 39 $ in this example) followed by 
$ h \simeq 2/[3 \; \tau ] $ in the  matter dominated era.}
\label{h}
\end{figure}

\begin{figure}[p]
\begin{turn}{-90}
\centering
\psfrag{"fi.dat"}{$ \chi $ vs. time } 
\psfrag{"fip.dat"}{$ \dot \chi $ vs. time }
\includegraphics[width=10cm,height=14cm]{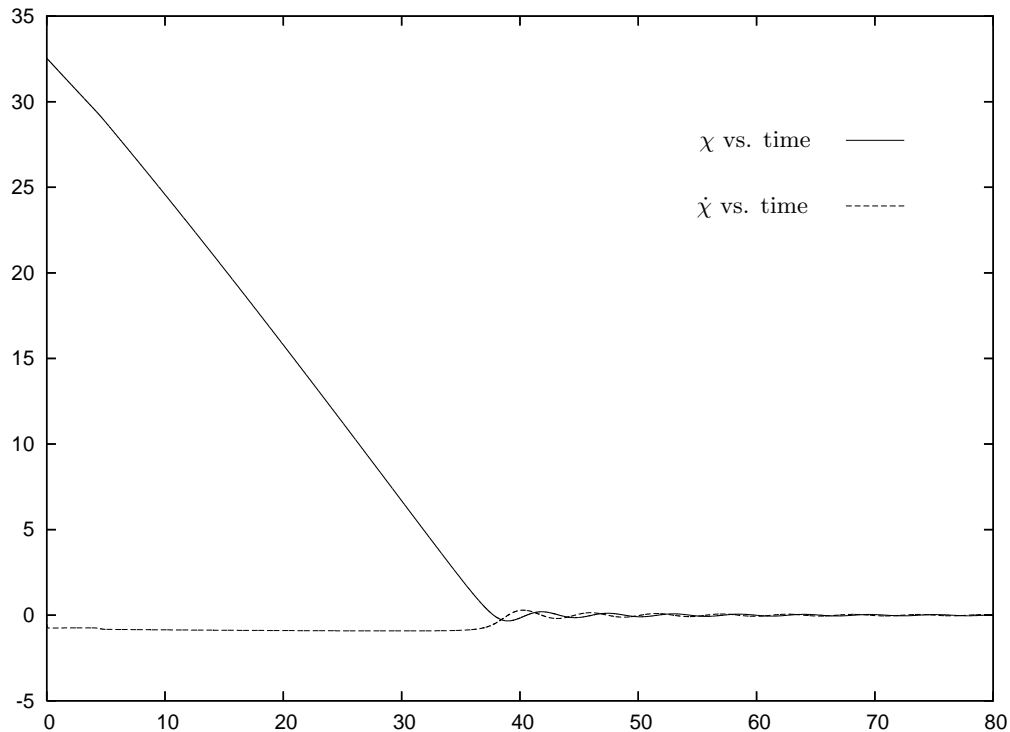}
\end{turn}
\caption{Hybrid inflation. The inflaton field $ \chi $ and its time derivative
as a function of time. Same parameters as in fig. \ref{a}.
$ |\dot \chi | \ll | \chi | $ during the slow-roll inflationary stage.
Inflation stops when $  \chi \sim {\dot \chi} \sim 0 $ 
 (at $ \tau \simeq 39 $ in this example).}
\label{fi}
\end{figure}

\begin{figure}[p]
\begin{turn}{-90}
\centering
\psfrag{"s.dat"}{$ \sigma $ vs. time } 
\psfrag{"sp.dat"}{$ \dot \sigma $ vs. time }
\includegraphics[width=10cm,height=14cm]{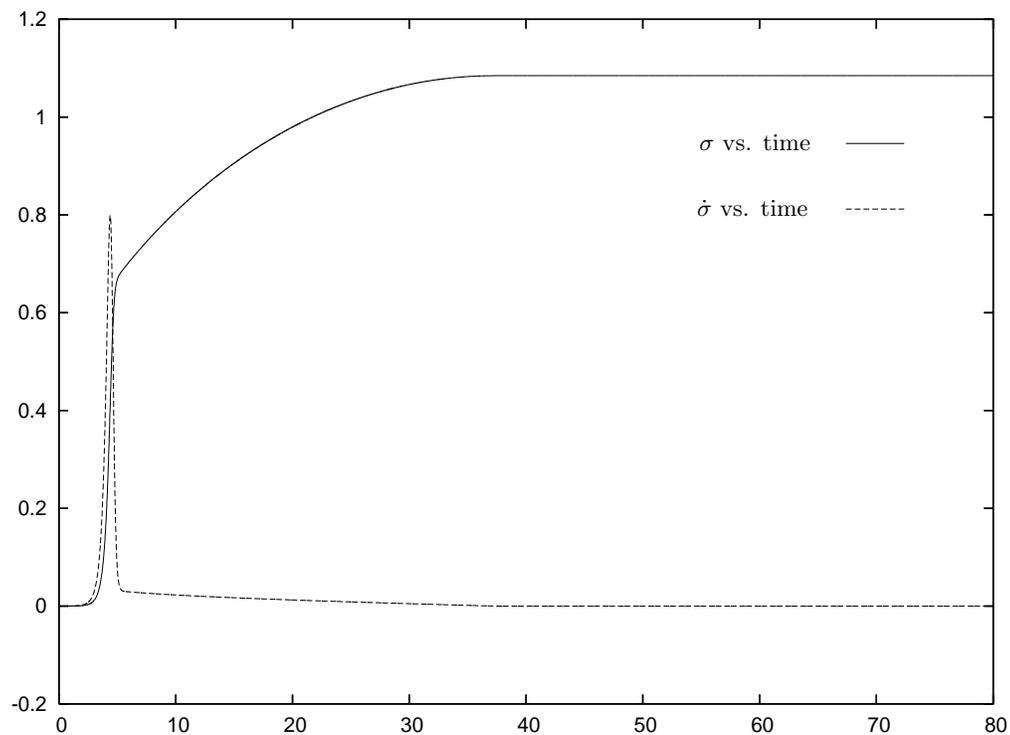}
\end{turn}
\caption{Hybrid inflation. The field sigma $ \sigma $ and its time derivative
as a function of time for hybrid inflation. Same parameters as in fig. \ref{a}.
The fields  $ \sigma $ and  $ \dot\sigma $ start with small values and grow 
exponentially fast when $ m_{\sigma}^2 < 0 $ [eq.(\ref{masefch})]
(at $ \tau \simeq 4 $ in this example). }
\label{si}
\end{figure}

\begin{figure}[p]
\begin{turn}{-90}
\centering
\psfrag{"pe.dat"}{pressure/energy density vs. time } 
\includegraphics[width=10cm,height=14cm]{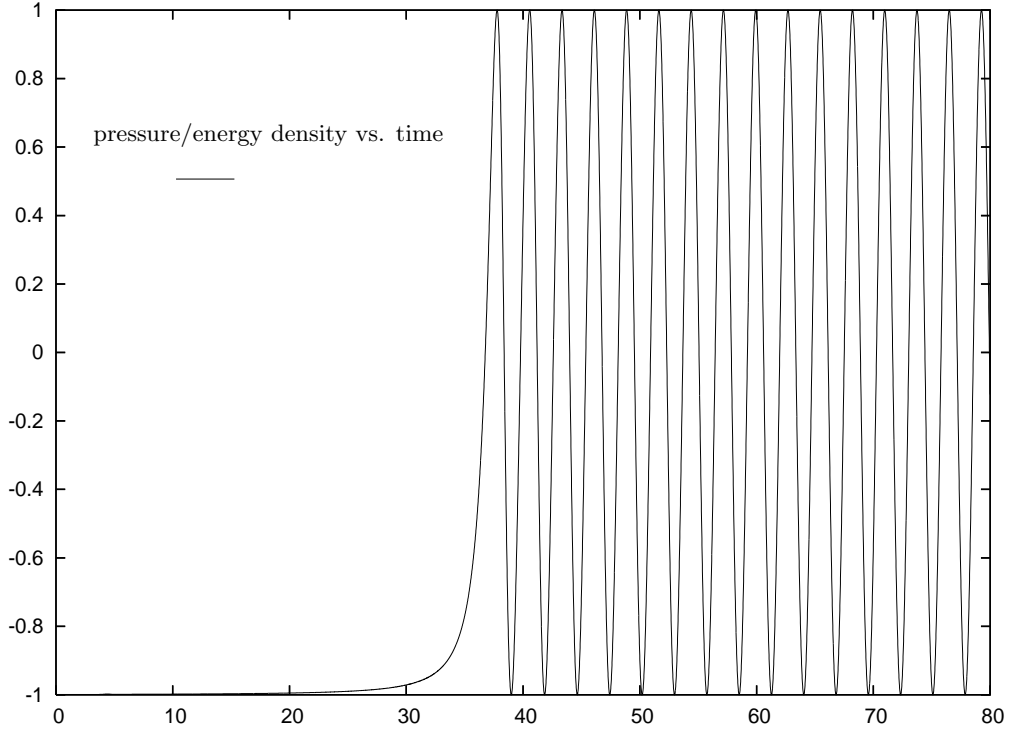}
\end{turn}
\caption{Hybrid inflation.The equation of state pressure/energy density
as a function of time. Same parameters as in fig. \ref{a}.
The equation of state clearly shows the two stages:
pressure/energy $ = - 1 $ during inflation followed by oscillations with zero average
pressure in the matter dominated era.}
\label{pe}
\end{figure}

\begin{figure}[p]
\begin{turn}{-90}
\centering
\psfrag{"nsr11.dat"}{${\hat \chi}(0)=0.01$}
\psfrag{"nsr9.dat"}{$ {\hat \chi}(0)=0.05 $}
\psfrag{"nsr1.dat"}{$ {\hat \chi}(0)=0.1 $} 
\psfrag{"nsr5.dat"}{$ {\hat \chi}(0)=0.2 $}
\psfrag{"nsr8.dat"}{$ {\hat \chi}(0)=0.3 $}
\psfrag{"nsr4.dat"}{$ {\hat \chi}(0)=0.4 $} 
\includegraphics[width=10cm,height=14cm]{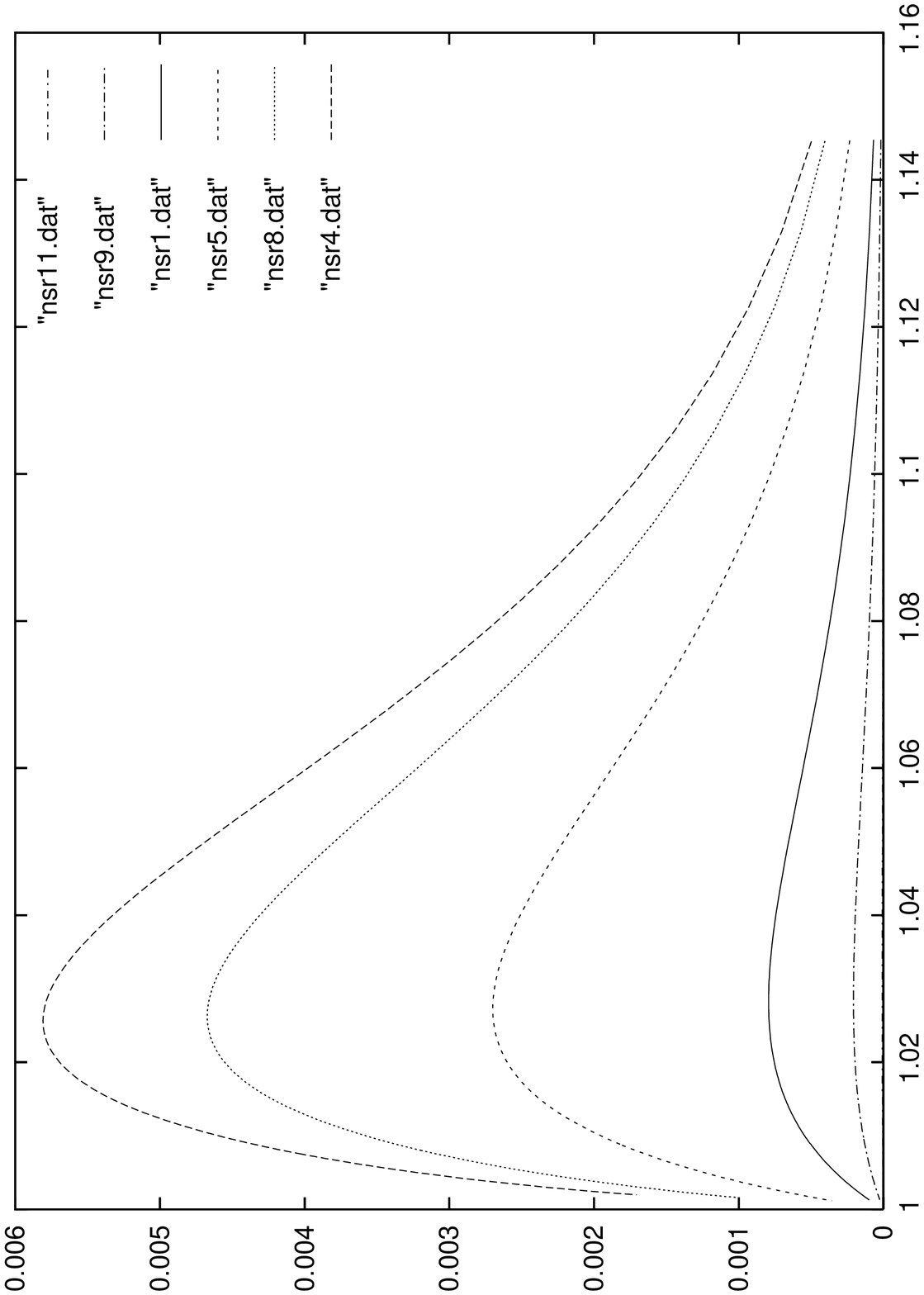}
\end{turn}
\caption{Hybrid inflation. The ratio $ r $ vs. $ n_s $ for $ \mu^2 =
0.05  \; \Lambda  <  \mu^2_{crit} \; , \; g^2 = \frac 14 $ and 
$ 0.01 \; \sqrt{\Lambda}\leq \chi(0)  
\leq 0.4 \; \sqrt{\Lambda} $. 
Notice that here $ r $ has a maximun as a function of $ n_s $.
In addition, $ n_s > 1 $ for all values of $ \Lambda $ and $ {\hat \chi}(0) $
since  $ \mu^2 < \mu^2_{crit} \simeq 0.13 \; \Lambda$. $ r $ increases with 
${\hat \chi}(0)$ for this range of ${\hat \chi}(0)$.}
\label{nsr005A}
\end{figure}

\section{Inflaton Dynamics in Hybrid Inflation}

In figs. \ref{a}-\ref{pe} we display the numerical solution of the 
equations of motion (\ref{ecmovh}) as functions of time for 
$ \Lambda = 4 \; N = 200 \; , \; \mu^2 = 1.7 \; \Lambda, \; 
\phi(0) = 2.3 \; \sqrt{\Lambda} $. We see first a stage of slow-roll
quasi-de Sitter inflation till $ \tau \simeq 39 $ in this example. 
Namely, $ | {\dot \phi} | \ll | \phi | $ and $ h $ are practically constant
during this lapse. In this slow-roll stage the equations of motion 
(\ref{ecmslo}) can be integrated in close form with the solution
\be \label{solslo}
\sqrt{\frac2{3 \; \Lambda}} (\tau - \tau_0) = {\rm Arg~ tanh}\frac1{\sqrt{
1 + \frac{\chi^2}{\Lambda} }} - \sqrt{1 + \frac{\chi^2}{\Lambda} } \; ,
\ee
which defines $ \chi = \chi(\tau) $ and where $ \tau_0 $ is an integration
constant. Notice that $ \chi(\tau) $ is a monotonically decreasing function
of time since $  {\dot \chi} = -\chi/[3 \; h] < 0 $ [eq.(\ref{ecmslo})].

When $ \chi \ll  \Lambda $, eq.(\ref{solslo})
approximates by
$$
\chi(\tau) \simeq \chi_0 \; e^{-\sqrt{\frac2{3 \; \Lambda}} \; \tau} \; ,
\quad  \chi \ll  \Lambda \quad \; ,
$$
while in the opposite limit $ \chi \gg \Lambda $
from eq.(\ref{solslo})  we have,
$$
\chi(\tau) \simeq \chi_1 - \sqrt{\frac23} \; \tau  \; ,
\quad  \chi \gg  \Lambda \quad \; .
$$
Here, $ \chi_0 $ and $ \chi_1 $ are integration constants.

\medskip

We have verified that eq.(\ref{solslo}) as well as eq.(\ref{Nhib}) provide
an excellent approximation to the numerical solution of eqs. (\ref{fihib}). 

The number of efolds during inflation is about
$280$ in the example depicted in figs. \ref{a}-\ref{pe}, 
larger than the required minimun of about $60$ efolds.
This stage is followed by a matter dominated era. 
We choose a very small initial  amplitude for the 
sigma field and its time derivative. The sigma field stays very small till
its effective mass square [eq.(\ref{masefe})] becomes negative and
spinodal unstabilities show up.
At this moment, ($ \tau \simeq 4 $ in this example) the sigma field as well as
its time derivative start to increase exponentially fast till the growth
of the non-linear term $ +\frac{\mu^4}{2 \; \Lambda} \; \sigma^2 $ 
in the last equation in eq.(\ref{ecmovh}) shuts off the unstabilities.

Inflation stops at the moment when both $ \chi $ and $ \sigma $ are comparable 
with $ \dot \chi $ and  $ \dot \sigma $ ($ \tau \simeq 39 $ in this example). 
At this time, both $ \chi $ and 
$ \sigma $ are very close to their vaccum values $ \chi_{vac} = 0 $ and
$ \sigma_{vac} = \frac{\sqrt{2 \; \Lambda}}{\mu} $. 
That is, when the kinetic terms become relevant, the energy is no more 
dominated by the vaccum energy. At the same time, the slow roll 
approximation ceases to be valid. 

The time when the effective mass of the field $ \sigma $ 
[see eq.(\ref{masefe})]
\be \label{masefch}
m_{\sigma}^2 = m^2 \; ( g^2 \; \chi^2 - \mu^2)
\ee
becomes negative and $ \sigma $ starts to grow depends on the
values of $ \mu^2 $ and $ g^2 \; \chi^2(0) $. For low values of 
$ \mu^2 $: (typically for $ \mu^2 < 0.08 \; \Lambda $ when 
$ \chi^2(0) < \Lambda $, and $ \mu^2 < 0.2 \; \Lambda $ when
$ \chi^2(0) < 2. \; \Lambda $), the field  $ \sigma $ starts to grow
close to the end of inflation. On the contrary, for higher values of 
$ \mu^2 $ (typically for $ \mu^2 > 0.08 \; \Lambda $ when
$ \chi^2(0) < \Lambda $, and  $ \mu^2 > 0.23 \; \Lambda $ when 
$ \chi^2(0) < 2. \; \Lambda $), the field  $ \sigma $ starts to grow
well before the end of inflation.  This is explained by the fact
that the scale of time variation of $ \sigma $ goes as
$ \mu^{-1} $;  $ \sigma $ evolves slowly for small $ \mu $ and
fastly for large $ \mu $.

\medskip

$ \dot \sigma $ exhibits a peak
around the point where $ m_{\sigma}^2 $ changes sign and then returns 
to a very small value while $ \sigma $ slowly approaches its vaccum
value. This evolution is depicted in fig. \ref{si}.

\medskip

In the example depicted in figs. \ref{a}-\ref{pe} the effective mass square 
[eq.(\ref{masefch})] of the $ \sigma $ field changes sign at $ \tau \sim 4 $ 
well before $ \chi $ reaches its vaccum value (zero) and inflation ends. 
This follows from the choice of a large value for $ \mu^2 $  in figs. 
\ref{a}-\ref{pe}. For smaller values of $ \mu^2 $, $ m_{\sigma}^2 $  
[eq.(\ref{masefch})] flips its sign later when the inflaton $ \chi $ is 
much smaller. 

\medskip

In order to compute the observables  $ n_s , \; r $ and $ d n_s /d \ln k $ 
from eq.(\ref{indhi}) we need the value of the inflation field
$ \chi $ at $50$ efolds before the end of inflation.
We thus integrated numerically eqs.(\ref{ecmovh}) till the end of
inflation and then extracted the value of $ \chi $ at $50$ efolds before.
We define the end of inflation as the point where the ratio
pressure over energy reaches $10\%$. This gives $ \tau \equiv \tau_{end}
\simeq 34 $ for the example in fig. \ref{pe}.

\medskip

At $ \Lambda = 0 $ hybrid inflation becomes chaotic inflation with
the monomial potential $ \frac12  \; \chi^2 $. We want to stress that
only {\bf at} $ \Lambda = 0 $ hybrid inflation becomes chaotic inflation.
For any  value of $ \Lambda > 0 $ (even very small) 
the features of hybrid inflation remain. The time $ \tau_{end} $
gets longer and longer for  $ \Lambda \to 0^+ $.

\section{Spectral index $n_s$, ratio $r$ and running index $ \frac{d n_s}{d \ln k} $
in Hybrid Inflation}

We see from eqs.(\ref{nshib}) that the field $ \chi $ naturally scales
as $ \sqrt{\Lambda} $. It is then convenient to introduce the rescaled
field and the rescaled mass
\be
{\hat \chi} \equiv \frac{\chi}{\sqrt{\Lambda}} \quad , \quad 
{\hat \mu}^2 \equiv  \frac{\mu^2}{\Lambda} \; .
\ee
Then, eqs.(\ref{nshib}) take the form
\bea\label{indhiR}
&& w(\chi,0) = \frac{\Lambda}2 \, ( {\hat \chi}^2 + 1) \quad , \cr \cr
&& |{\Delta}_{k\;ad}^{(S)}|^2 = \frac{N^2 \; \Lambda^2}{96 \, \pi^2 } 
\left( \frac{M}{M_{Pl}} 
\right)^4 \; \frac{( {\hat \chi}^2 + 1)^3}{ {\hat \chi}^2} \quad ,  \quad 
r = \frac{32}{N\; \Lambda} \; 
\frac{ {\hat \chi}^2}{({\hat \chi}^2 + 1)^2}  \; ,  \\ \cr
&&n_s = 1 + \frac4{N \; \Lambda} \; 
\frac{1 - 2 \; {\hat \chi}^2}{({\hat \chi}^2 + 1)^2}
\quad ,  \quad \frac{d n_s}{d \ln k}= \frac{32}{N^2\; \Lambda^2} \; 
\frac{{\hat \chi}^2(2 -{\hat \chi}^2)}{({\hat \chi}^2 + 1)^4} 
\label{nshibR} \; . 
\eea
Notice that $ (n_s - 1) $ may have either sign according to eq.(\ref{nshibR}).
Hybrid inflation is usually associated with red tilted spectrum ($ n_s > 1 $).
However, both regimes, $ n_s > 1 $ {\bf and} $ n_s < 1 $ are
realized by hybrid inflation.

\medskip

\begin{figure}[p]
\begin{turn}{-90}
\centering
\psfrag{"nsr3.dat"}{${\hat \chi}(0)=2.4$}
\psfrag{"nsr6.dat"}{${\hat \chi}(0)=0.8$}
\psfrag{"nsr7.dat"}{${\hat \chi}(0)=0.5$}
\psfrag{"nsr10.dat"}{${\hat \chi}(0)=1.0$}
\psfrag{"nsr.dat"}{$  {\hat \chi}(0)=0.6 $}
\includegraphics[width=10cm,height=14cm]{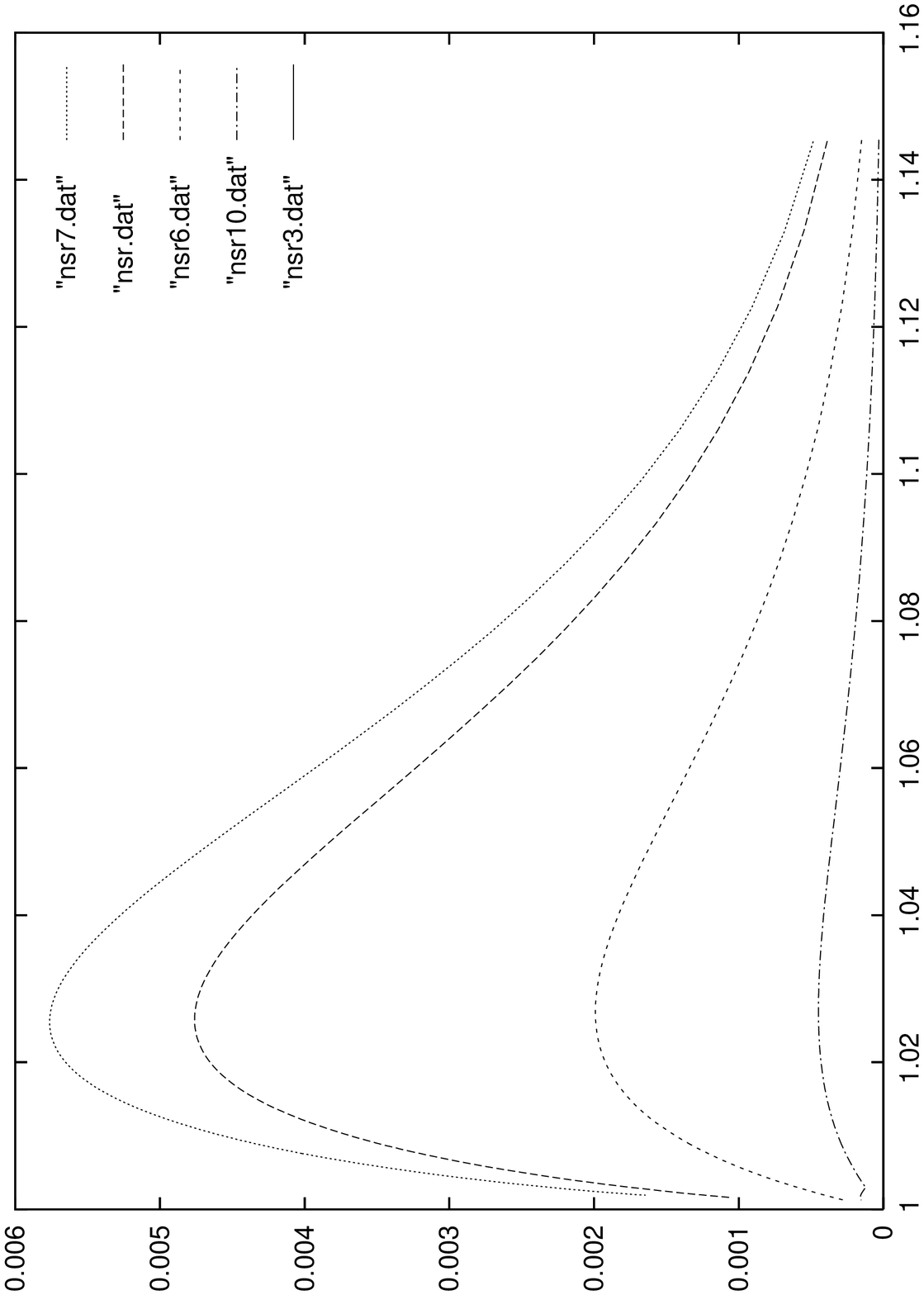}
\end{turn}
\caption{Hybrid inflation. The ratio $ r $ vs. $ n_s $ for $ \mu^2 =
0.05  \; \Lambda   <  \mu^2_{crit} \; , \; g^2 = \frac 14 $ and 
$ 0.5 \; \sqrt{\Lambda}\leq \chi(0)  
\leq 2.4 \; \sqrt{\Lambda} $. 
Notice that here $ r $ has a maximun as a function of $ n_s $.
In addition, $ n_s > 1 $ for all values of $ \Lambda $ and $ \chi(0) $ since  
$ \mu^2 < \mu^2_{crit} \simeq 0.13 \; \Lambda$. $ r $ decreases with 
${\hat \chi}(0)$ for this range of ${\hat \chi}(0)$.}
\label{nsr005B}
\end{figure}

\begin{figure}[p]
\begin{turn}{-90}
\centering
\psfrag{"nsr4.dat"}{${\hat \chi}(0)=0.7$}
\psfrag{"nsr5.dat"}{${\hat \chi}(0)=0.9$}
\psfrag{"nsr6.dat"}{${\hat \chi}(0)=1.1$}
\psfrag{"nsr7.dat"}{${\hat \chi}(0)=1.3$}
\psfrag{"nsr8.dat"}{${\hat \chi}(0)=1.5$}
\psfrag{"nsr9.dat"}{${\hat \chi}(0)=1.7$}
\psfrag{"nsr10.dat"}{${\hat \chi}(0)=1.9$}
\psfrag{"nsr11.dat"}{${\hat \chi}(0)=2.1$}
\psfrag{"nsr12.dat"}{${\hat \chi}(0)=2.3$}
\includegraphics[width=10cm,height=14cm]{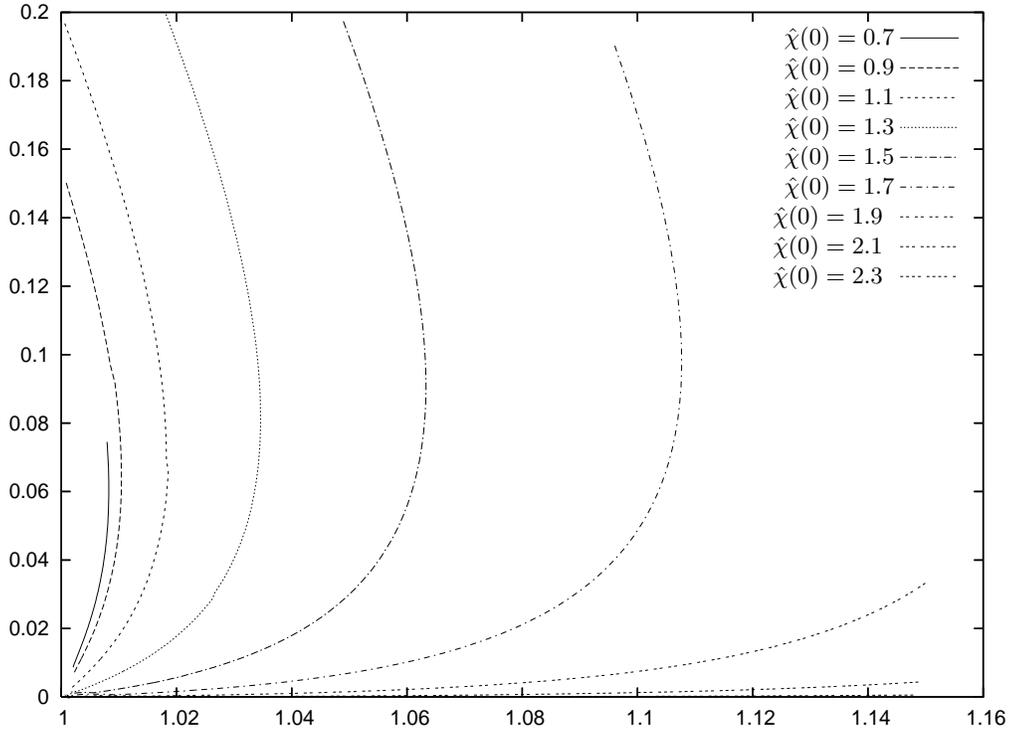}
\end{turn}
\caption{Hybrid inflation. The ratio $ r $ vs. $ n_s $ for $ \mu^2 =
0.13 \; \Lambda \simeq \mu^2_{crit} \; , \; g^2 = \frac 14 $ and 
$ 0.7 \; \sqrt{\Lambda}\leq \chi(0) \leq 2.3 \; \sqrt{\Lambda} $. 
Notice that here $ n_s > 1 $ for all values of $ \Lambda $ and $ \chi(0) $.}
\label{nsrh}
\end{figure}

\begin{figure}[p]
\begin{turn}{-90}
\centering
\psfrag{"nsrun4.dat"}{${\hat \chi}(0)=0.7$}
\psfrag{"nsrun5.dat"}{${\hat \chi}(0)=0.9$}
\psfrag{"nsrun6.dat"}{${\hat \chi}(0)=1.1$}
\psfrag{"nsrun7.dat"}{${\hat \chi}(0)=1.3$}
\psfrag{"nsrun8.dat"}{${\hat \chi}(0)=1.5$}
\psfrag{"nsrun9.dat"}{${\hat \chi}(0)=1.7$}
\psfrag{"nsrun10.dat"}{${\hat \chi}(0)=1.9$}
\psfrag{"nsrun11.dat"}{${\hat \chi}(0)=2.1$}
\psfrag{"nsrun12.dat"}{${\hat \chi}(0)=2.3$}
\includegraphics[width=10cm,height=14cm]{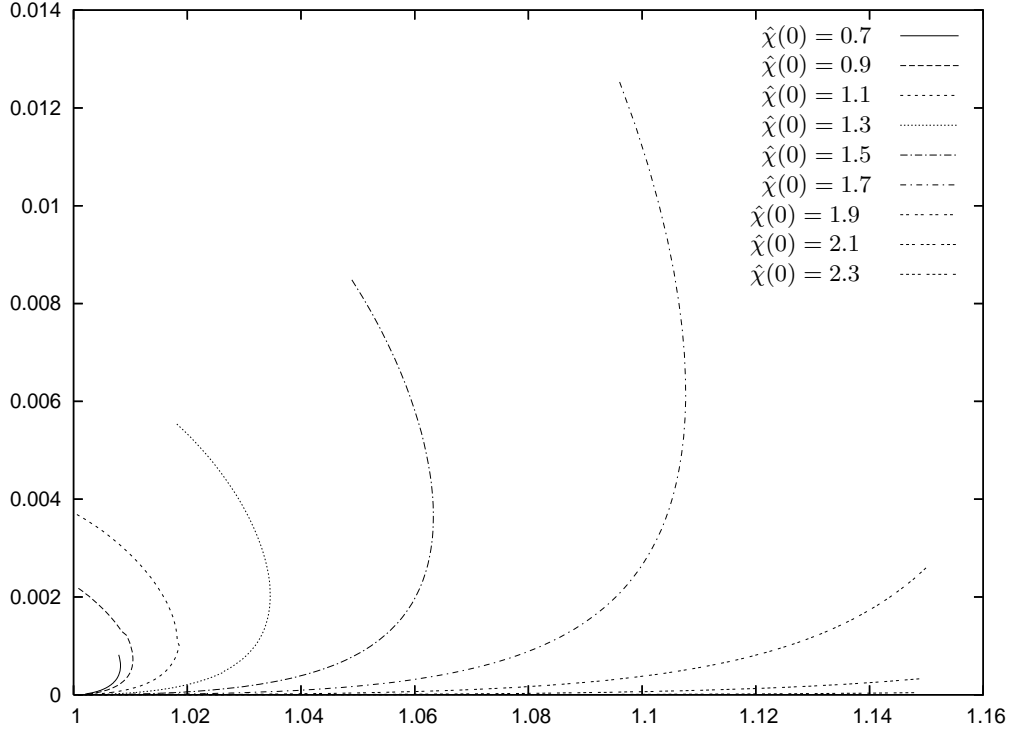}
\end{turn}
\caption{Hybrid inflation. The running $ d n_s/ d \ln k $ vs. $ n_s $ for 
$ \mu^2 = 0.13 \; \Lambda \simeq \mu^2_{crit} \; , \; g^2 = \frac 14 $ and 
$ 0.7 \; \sqrt{\Lambda}\leq \chi(0)  \leq 2.3 \; \sqrt{\Lambda} $. 
Notice that here $  d n_s/ d \ln k > 0 $ for all values of $ \Lambda $ and 
$ \chi(0) $. It exhibits a shape similar to $ r $ vs. $ n_s $ 
[see fig. \ref{nsr005B}].}
\label{nsrunh}
\end{figure}

\begin{figure}[p]
\begin{turn}{-90}
\centering
\psfrag{"ns7.dat"}{${\hat \chi}(0)=2.7$}
\psfrag{"ns10.dat"}{${\hat \chi}(0)=3.3$}
\psfrag{"ns12.dat"}{${\hat \chi}(0)=3.7$}
\psfrag{"ns14.dat"}{${\hat \chi}(0)=4.1$}
\psfrag{"ns16.dat"}{${\hat \chi}(0)=4.5$}
\psfrag{"ns17.dat"}{${\hat \chi}(0)=4.7$}
\psfrag{"ns19c.dat"}{${\hat \chi}(0)=5.1$}
\psfrag{"ns20c.dat"}{${\hat \chi}(0)=5.3$}
\psfrag{"ns21c.dat"}{${\hat \chi}(0)=5.7$}
\includegraphics[width=10cm,height=14cm]{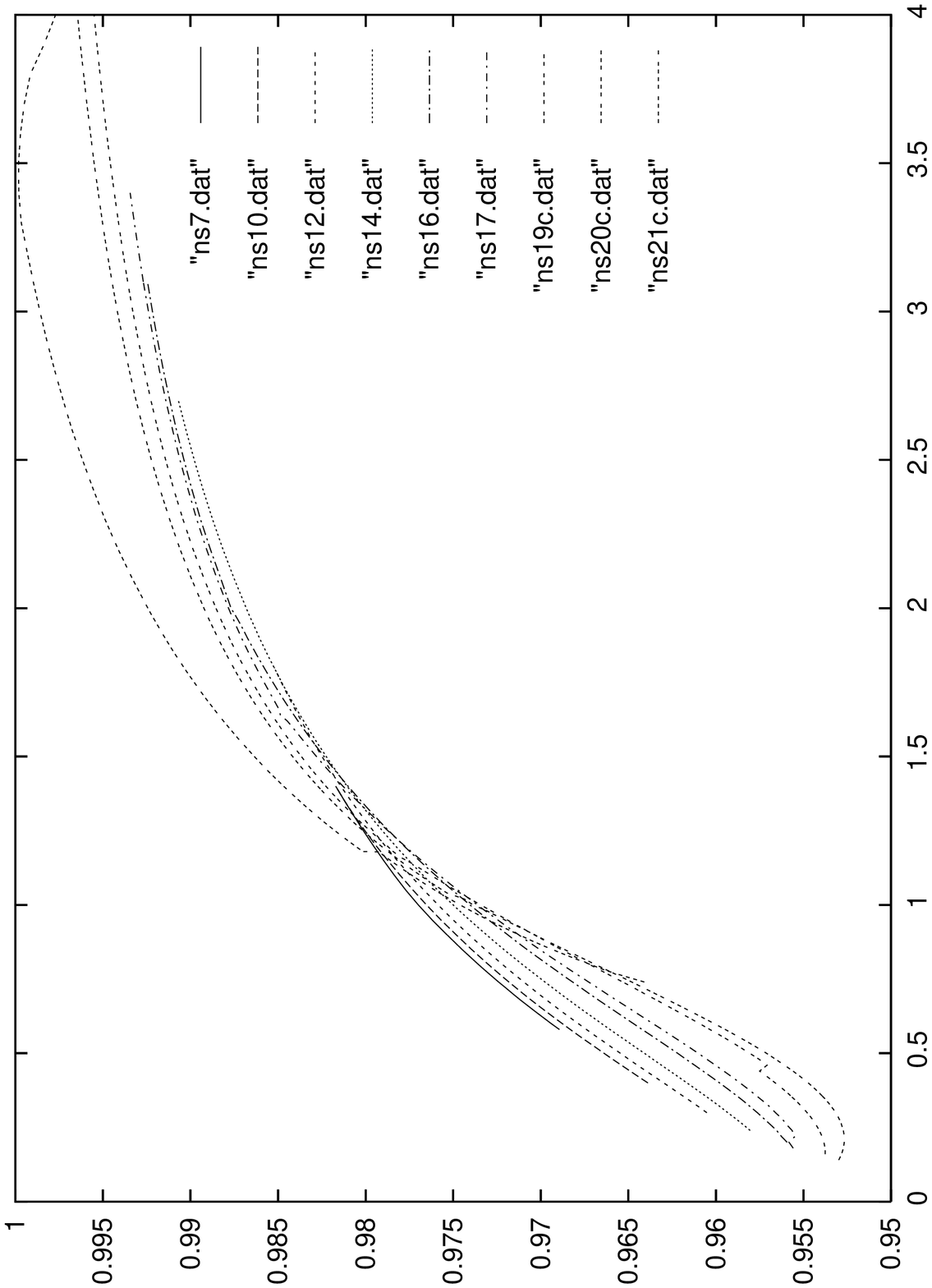}
\end{turn}
\caption{Hybrid inflation. The index $ n_s  $ vs. $ \Lambda $ for $ \mu^2 =
1.7 \; \Lambda > \mu^2_{crit} , \; g^2 = \frac 14 $ and 
$ 2.7 \; \sqrt{\Lambda}\leq \chi(0) \leq 5.7 \; \sqrt{\Lambda} \simeq 
\chi(0)_{crit} $. Notice that here $ n_s < 1 $ for all values of $ \Lambda $ 
and this range of $ \chi(0) $.}
\label{nsY}
\end{figure}

Whether $ n_s > 1 $ or $ n_s < 1 $ for a given set of parameters
$ \Lambda,  \; \mu^2 , \; g $ and the initial conditions
is a {\bf dynamical} question that can only be answered after
evolving the fields till the end of inflation according to
eqs.(\ref{ecmovh}). As we see in eq.(\ref{nshib}) the question is whether twice
$ \chi^2 $ at horizon exit is larger or smaller than $ \Lambda $. Recall that
horizon exit happens about $50$ efolds before the end of inflation and that
$ \chi^2 $ monotonically decreases during inflation. Even if initially
$ 2 \; \chi^2 > \Lambda $, it can be very well that 
$ 2 \; \chi^2 < \Lambda $ at horizon exit. 
This depends on  how many total efolds 
$ N_T \geq 60 $ we have; since horizon exit happens $ (N_T - 50) \geq 10 $ 
efolds after the beginning of inflation, the larger is  $ (N_T - 50) $
the smaller can be $ \chi^2 $ at horizon exit. 

We vary the parameters $ \Lambda,  \; \mu^2 $ and the initial conditions
always keeping the total number of efolds $ N_T $ during inflation larger
or equal to $60$. We keep $ g^2 = \frac14 $ since this parameter is less
relevant than the others. We explored the parameters region where $ r < 0.2 $
and $ 0.95 < n_s < 1.15 $. 

\subsection{Red tilted and blue tilted regimes in Hybrid Inflation}

Extended numerical investigation showed that there exists a {\bf critical 
value} of $ \mu^2 , \; \mu^2_{crit} \simeq 0.13 \;  \Lambda  $ such that 
$ n_s > 1 $  provided $ \mu^2 < \mu^2_{crit} $.

\medskip

For $ \mu^2 > \mu^2_{crit} $ we find both regimes,  $ n_s > 1 $ {\bf and} 
$ n_s < 1 $. This property is valid for all initial values of the inflaton
compatible with the restrictions $ N_T \geq 60 $ and one-inflaton fluctuations.
Otherwise, if the field $ \sigma $ is relevant at horizon exit we should also
include its contribution to the density fluctuations. Such calculation
is beyond the scope of the present work where we concentrate on single inflaton
fluctuations.

The larger is $ \mu^2 $, the earlier inflation ends, the earlier horizon
exit happens and the large is $ \chi $ at horizon exit. 
That is, increasing $ \mu^2 $ decreases $ n_s $. This explains why we 
necessarily find $ n_s < 1 $ for $ \mu^2 > \mu^2_{crit} $.

\medskip

For $ \mu^2 > \mu^2_{crit} $ we find that $ n_s > 1 $ for
$ \chi(0) >  \chi(0)_{crit} $. That is, increasing $ \chi(0) $, increases 
$ n_s $. This is so because the larger is $ \chi(0) $, the larger is $ N_T $ since 
$ N_T \sim  \chi(0)^2 $  [see eqs.(\ref{ecmslo}) and (\ref{Nhib})]. 
Then, the larger is $ (N_T - 50) $ the smaller is $ \chi $ at horizon exit 
and the larger is $ n_s $. 

\medskip

In all cases, (both $ \mu^2 > \mu^2_{crit} $ and  $ \mu^2 < \mu^2_{crit} $)
for $ \Lambda \to \infty $ we always find 
$ n_s \to 1, \; r \to 0 $ and $ \frac{d n_s}{d \ln k}  \to 0 $.

\medskip

\begin{figure}[p]
\begin{turn}{-90}
\centering
\psfrag{"ns21.dat"}{${\hat \chi}(0)=5.7$}
\psfrag{"ns21a.dat"}{${\hat \chi}(0)=5.9$}
\psfrag{"ns22.dat"}{${\hat \chi}(0)=6.1$}
\psfrag{"ns23.dat"}{${\hat \chi}(0)=6.5$}
\psfrag{"ns24.dat"}{${\hat \chi}(0)=6.9$}
\psfrag{"ns25.dat"}{${\hat \chi}(0)=7.3$}
\psfrag{"ns26.dat"}{${\hat \chi}(0)=7.7$}
\psfrag{"ns27.dat"}{${\hat \chi}(0)=8.1$}
\psfrag{"ns28.dat"}{${\hat \chi}(0)=8.7$}
\psfrag{"ns30.dat"}{${\hat \chi}(0)=10.7$}
\includegraphics[width=10cm,height=14cm]{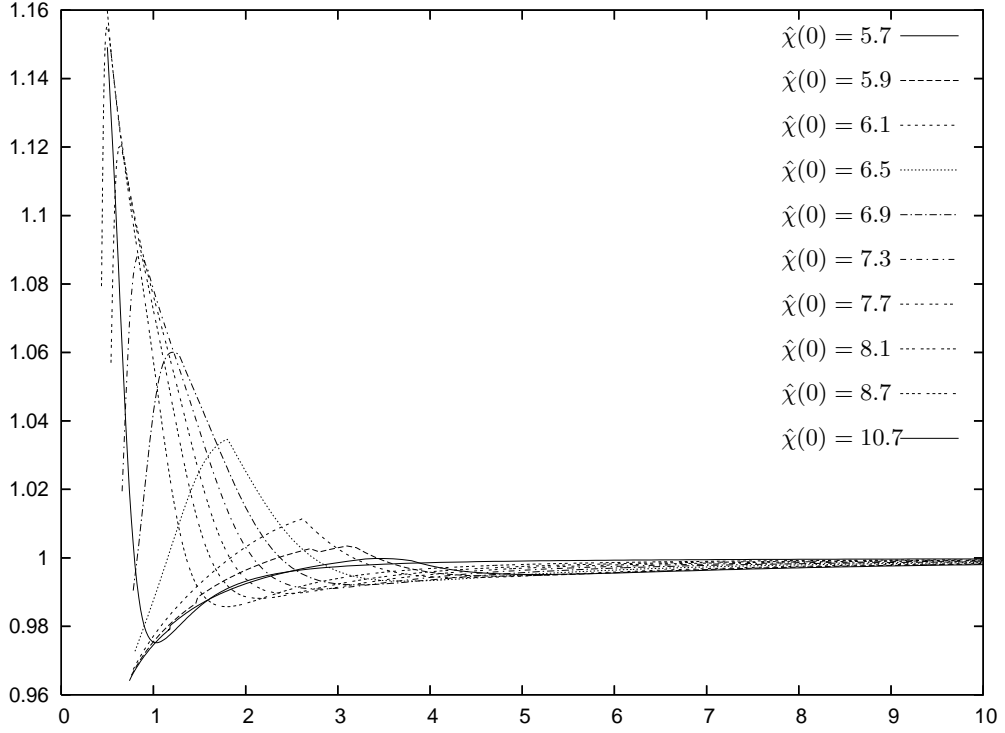}
\end{turn}
\caption{Hybrid inflation. The index $ n_s $ vs. $ \Lambda $ for $ \mu^2 =
1.7 \; \Lambda  > \mu^2_{crit} , \; g^2 = \frac 14 $  and 
$ \chi(0)_{crit} \simeq 5.7 \; \sqrt{\Lambda}\leq \chi(0) \leq 10.7 
\; \sqrt{\Lambda} $. Notice that here we have both $ n_s > 1 $ and  
$ n_s < 1 $ depending on the values of $ \Lambda $ and $ \hat \chi(0) $.
All curves reach asymptotically $ n_s = 1 $ for $ \Lambda \to \infty $.}
\label{nsA}
\end{figure}

Figs. \ref{nsr005A}-\ref{nsrunA} show the observables 
$ n_s , \; r $ and the running index 
$ d n_s /d \ln k $ as  functions of $ \Lambda $ and $ n_s $ for 
$  \mu^2 = 0.05 \; \Lambda , \;  \mu^2 = 0.13 \; \Lambda $ and 
$ \mu^2 = 1.7  \;  \Lambda $. A complete picture for 
hybrid inflation emerges covering {\bf two} different, blue tilted and 
red tilted,  regimes. We find that for all the observables, the shape of 
the curves depends crucially on the mass parameter $ {\hat \mu}^2 $ of the 
$\sigma$ field and the (rescaled) initial amplitude 
$ {\hat \chi}(0) $ of the inflaton field.

We find three regimes according to the value of $ {\hat \mu}^2 $:

\begin{itemize}

\item $ {\hat \mu}^2 < 0.075 $. Here we always have $ n_s > 1 $ and $ r $ 
has one maximun as a function of $ n_s $ (or $ \Lambda $). 
 $ n_s $ monotonically decreases with $ \Lambda $.
Figs. \ref{nsr005A} and \ref{nsr005B} show $ r $ vs. $ n_s $ for
$  \mu^2 = 0.05 \; \Lambda $ and various values of $ {\hat \chi}(0) $.
$ r $ displays a maximun as a function of $ n_s $.
In addition, $ r $ grows with  $  {\hat\chi}(0) $ for  $  {\hat\chi}(0) < 0.5 $
while it decreases with  $ {\hat\chi}(0) $ for  $  {\hat\chi}(0) > 0.5 $.
The running  $ \frac{d n_s}{d \ln k} $ behaviour is qualitatively
similar to the behaviour of $ r $ above described. 
The running $ \frac{d n_s}{d \ln k} $ is here positive and grows when 
$ n_s $ grows.

\item $ 0.075 < {\hat \mu}^2 < {\hat \mu}^2_{crit} \simeq 0.13 $. 
Here we always have $ n_s > 1 $ and $ r $ monotonically grows with $ \Lambda $.
$ n_s $ monotonically decreases with $ \Lambda $ for $ {\hat \mu}^2 < 0.1 $
while it exhibits a maximun as a function of $ \Lambda $ for  $ {\hat \mu}^2 
> 0.1 $. Figs. \ref{nsrh} and \ref{nsrunh} depict  $ r $ and
the running $ \frac{d n_s}{d \ln k} $  vs. $ n_s $, respectively,
for $ {\hat \mu}^2 = 0.13 \simeq {\hat \mu}^2_{crit} $. 
We see that the running  $ \frac{d n_s}{d \ln k} $ behaviour 
is qualitatively similar to the one of $ r $.

\item $ {\hat \mu}^2 >  0.13 \simeq {\hat \mu}^2_{crit} $.
Here $ n_s > 1 $ for  $ {\hat \chi}(0) > {\hat \chi}(0)_{crit} $
and  $ n_s < 1 $ for  $ {\hat \chi}(0) < {\hat \chi}(0)_{crit} $.

The value of $ {\hat \chi}(0)_{crit} $ grows with $ {\hat \mu}^2 $:
for $ {\hat \mu}^2 = 0.5 $,  we find  $ {\hat \chi}(0)_{crit} = 2.7$
and for $ {\hat \mu}^2 = 1.7 $, we find   $ {\hat \chi}(0)_{crit} = 5.8$.

For $ {\hat \chi}(0) < {\hat \chi}(0)_{crit} , \; 
n_s $ monotonically increases with $ \Lambda $ with values $ n_s < 1 $. 
For $ {\hat \chi}(0) > {\hat \chi}(0)_{crit} , \; n_s $ shows an absolute 
{\bf maximum}, which is always $ n_{s~max} > 1 $. The highest $n_s$ values 
concentrate and narrow in the small $ \Lambda $ region. 
It must be noticed that for each curve, [each $ {\hat \chi}(0) $],  
$ n_s $ can take values $ n_s >1 $ {\bf and}
$ n_s < 1 $: even if $ n_{s~max} > 1 , \; n_s $ can be below unit in the two 
sides of the curve, [see figs. \ref{nsY} and \ref{nsA}].

The value $ n_s = 1 $,  is reached  asymptotically for large $\Lambda$ from 
$ n_s < 1 $ for both $ {\hat \chi}(0) < {\hat \chi}(0)_{crit} $ and  
$ {\hat \chi}(0) > {\hat \chi}(0)_{crit} $. In addition, the value $ n_s = 1 $
with $ 0.2 > r > 0.04 $ is found for a variety of values of $ \Lambda $
and  $ {\hat \chi}(0) > {\hat \chi}(0)_{crit} $ as we see from figs. \ref{nsA}, 
\ref{rA} and \ref{nsrA}.

\end{itemize}

\subsection{The ratio $r$ in Hybrid Inflation}

\begin{figure}[p]
\begin{turn}{-90}
\centering
\psfrag{"r7.dat"}{${\hat \chi}(0)=2.7$}
\psfrag{"r10.dat"}{${\hat \chi}(0)=3.3$}
\psfrag{"r12.dat"}{${\hat \chi}(0)=3.7$}
\psfrag{"r14.dat"}{${\hat \chi}(0)=4.1$}
\psfrag{"r16.dat"}{${\hat \chi}(0)=4.5$}
\psfrag{"r17.dat"}{${\hat \chi}(0)=4.7$}
\psfrag{"r19c.dat"}{${\hat \chi}(0)=5.1$}
\psfrag{"r20c.dat"}{${\hat \chi}(0)=5.3$}
\psfrag{"r21c.dat"}{${\hat \chi}(0)=5.7$}
\includegraphics[width=10cm,height=14cm]{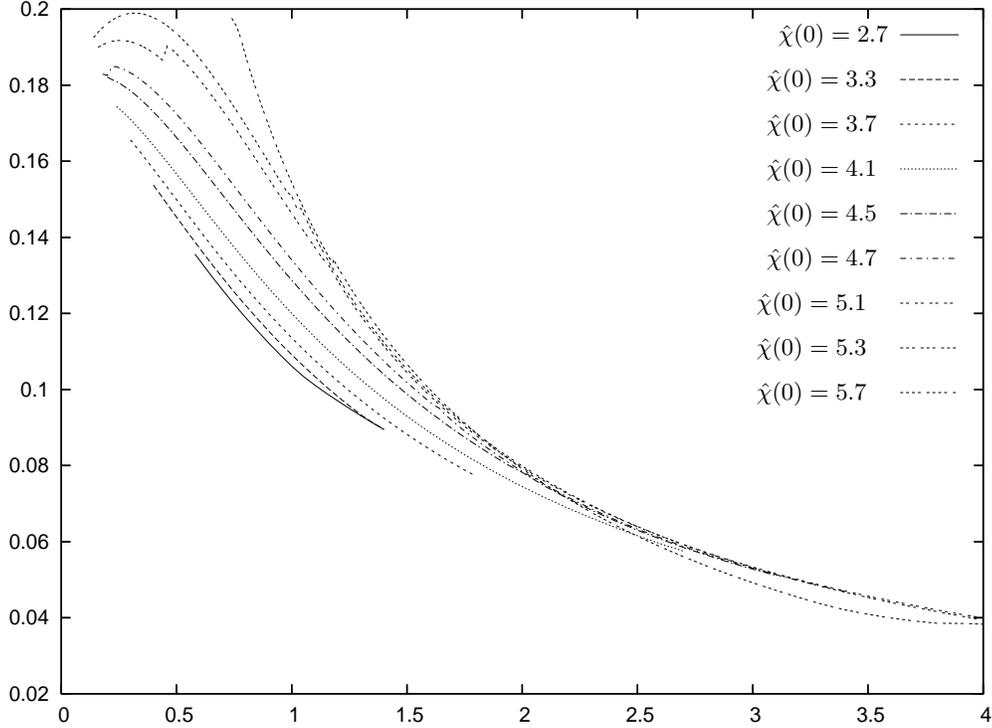}
\end{turn}
\caption{Hybrid inflation. The ratio $ r $ vs. $ \Lambda $ for $ \mu^2 =
1.7 \; \Lambda > \mu^2_{crit} , \; g^2 = \frac 14 $ and $ 2.7 \; 
\sqrt{\Lambda}\leq \chi(0) \leq 5.7 \; \sqrt{\Lambda}\simeq \chi(0)_{crit} $. 
$ r $ decreases monotonically with $ \Lambda $ in this regime 
$ \chi(0) < \chi(0)_{crit} $ and asymptotically vanishes for $ \Lambda \to \infty $.}
\label{rY}
\end{figure}

\begin{figure}[p]
\begin{turn}{-90}
\centering
\psfrag{"r21.dat"}{${\hat \chi}(0)=5.7$}
\psfrag{"r22.dat"}{${\hat \chi}(0)=6.1$}
\psfrag{"r23.dat"}{${\hat \chi}(0)=6.5$}
\psfrag{"r24.dat"}{${\hat \chi}(0)=6.9$}
\psfrag{"r25.dat"}{${\hat \chi}(0)=7.3$}
\psfrag{"r26.dat"}{${\hat \chi}(0)=7.7$}
\psfrag{"r27c.dat"}{${\hat \chi}(0)=8.1$}
\psfrag{"r28.dat"}{${\hat \chi}(0)=8.7$}
\psfrag{"r30.dat"}{${\hat \chi}(0)=10.7$}
\includegraphics[width=10cm,height=14cm]{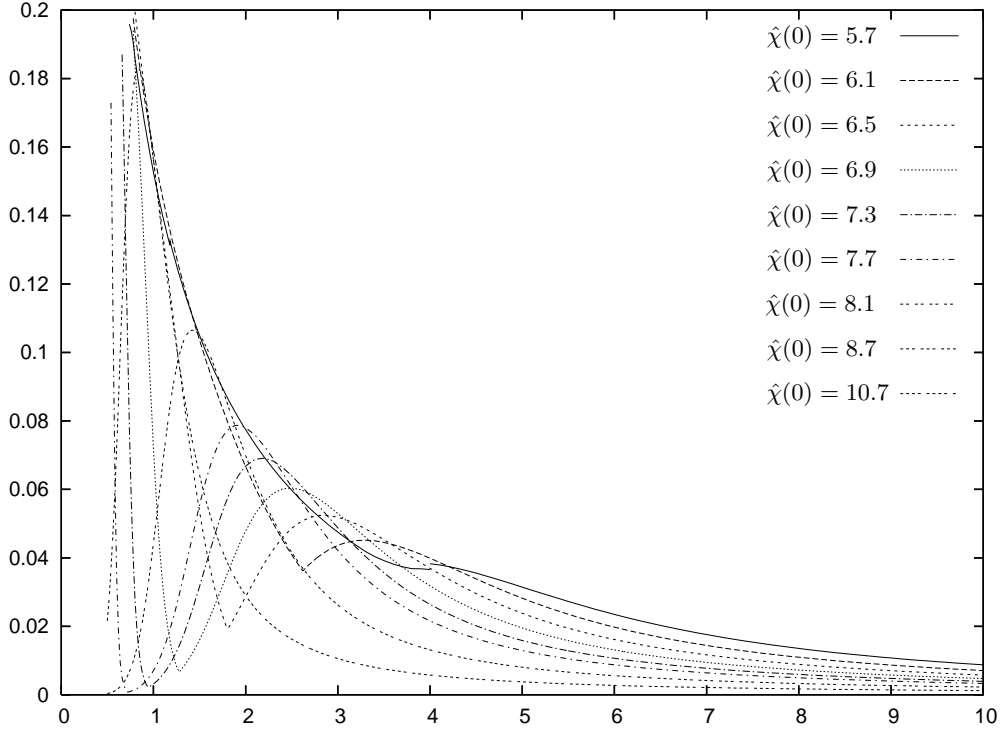}
\end{turn}
\caption{Hybrid inflation. The ratio $ r $ vs. $ \Lambda $ for $ \mu^2 =
1.7 \; \Lambda  > \mu^2_{crit} , \; g^2 = \frac14 $
and $ \chi(0)_{crit} \simeq  5.7 \; \sqrt{\Lambda}\leq \chi(0)  
\leq 10.7 \; \sqrt{\Lambda} . \;  r $ shows an oscillatory pattern in this
regime $ \chi(0)  \geq \chi(0)_{crit} $ and asymptotically vanishes for 
$ \Lambda \to \infty $ with no oscillations. The oscillations show up and concentrate
with increasing amplitude for small $ \Lambda $.}
\label{rA}
\end{figure}

The ratio $r$ in figs. \ref{rY} and \ref{rA} exhibits an  
{\bf oscillatory} pattern and two different regimes:

For  $ {\hat \chi}(0) < {\hat \chi}(0)_{crit} $, (fig. \ref{rY}) 
$ r $ decreases monotonically 
reaching very small values for large $\Lambda$. For  
$ {\hat \chi}(0) < {\hat \chi}(0)_{crit} $, $ r $ 
does not feature any oscillation. 

For $ {\hat \chi}(0) > {\hat \chi}(0)_{crit} , \; r $ decreases 
with $ \Lambda $, (fig. \ref{rA}) $ r $ has an absolute {\bf minimum} 
$ r_{min} $ and then grows till a {\bf maximum}, $ r_{max} $. The oscillations 
show up and concentrate with growing amplitude for small 
$ \Lambda $ for high $ {\hat \chi}(0) $,  $ r_{min} $ and  $ r_{max} $ 
shift towards the 
smaller $\Lambda$ with increasing $ {\hat \chi}(0) $;  $r_{min}$ decreases, and
 $r_{max}$ increases, for increasing $  {\hat \chi}(0)  $.
The convexity of the curve for small $ \Lambda $ increases
for decreasing $ {\hat \chi}(0) $.

For $ {\hat \chi}(0) > {\hat \chi}(0)_{crit} $,
each curve [each ${\hat \chi}(0)$] shows for $ r $ a 
 {\bf oscillatory} behavior with three clear parts: (1) the asymptotic part of 
monotonically decreasing $ r $ for large $ \Lambda $ at the right of 
$r_{max}$; (2) the increasing part at the left of $r_{max}$; (3) the sharp 
decreasing part for small $\Lambda$ at the left of $r_{min}$.
In the minima, $r_{min}$ can be extremely small for small 
$\Lambda$, which is a  {\bf new feature} in hybrid inflation.

In the asymptotic regime of large $\Lambda$, $r$ does not feature 
any oscillation. All curves [for all ${\hat \chi}(0) $] coalesce into $ r = 0 $
for $ \Lambda \to \infty $.

There are three distinct regimes: small $\Lambda$,  intermediate
$\Lambda$ and large $\Lambda$. 
The new oscillatory behavior for high and intermediate 
$ {\hat \chi}(0) > {\hat \chi}(0)_{crit} $
is in the region of small and intermediate 
$\Lambda$. The monotonically decreasing behavior for low 
${\hat \chi}(0) < {\hat \chi}(0)_{crit} $ 
is in the asymptotic region of large $\Lambda$.

\medskip

The highest values of $ r $ appear for small $\Lambda$ whatever be the 
hybrid regime; for such high values of $ r $ both regimes 
$ {\hat \chi}(0) < {\hat \chi}(0)_{crit} $ 
and $ {\hat \chi}(0) > {\hat \chi}(0)_{crit} $
superpose. From such high values, $r$ decreases  sharply till its minimun 
$r_{min}$ in the small $\Lambda$ region for 
${\hat \chi}(0) > {\hat \chi}(0)_{crit}$;
or $r$ decreases monotonically reaching asymptotically 
the large $ \Lambda $ regime for ${\hat \chi}(0) < {\hat \chi}(0)_{crit}$.
The low values of $ r $ for low $ \Lambda $ are a {\bf totally  new} 
feature in hybrid inflation.

\subsection{The running index $ d n_s /d \ln k $ in Hybrid Inflation}

The curves of the running index $ d n_s /d \ln k $ figs. \ref{nsrunh},
\ref{nsrunY} and \ref{nsrunA} show {\bf new} features in two different regimes.

\medskip

For $ {\hat \chi}(0) > {\hat \chi}(0)_{crit} $, the running index  
$ d n_s /d \ln k $ shows a similar shape as $r$ and is essentially 
positive. It oscillates with at least one {\bf maximum} 
and one or two {\bf minima} and three different components: (1) the 
asymptotic part of monotonically decreasing running with increasing 
$\Lambda$, at the right of the maximum, going 
to zero in this regime;  (2) the increasing running with $\Lambda$, 
which is a new feature in hybrid inflation, and (3) the sharp 
decreasing of the running till its minimum value for small $\Lambda$. 
The highest running appears
for small $\Lambda$ as in the known hybrid regime. The lower running 
values for small $\Lambda$, as well as the  {\bf oscillations} for small and 
intermediate $\Lambda$, are totally {\bf new}. \\

For low  $ {\hat \chi}(0) < {\hat \chi}(0)_{crit} $ the running index
does not exhibit any oscillation.
Both for $ {\hat \chi}(0) < {\hat \chi}(0)_{crit} $ and  
${\hat \chi}(0) > {\hat \chi}(0)_{crit}$
$ d n_s /d \ln k $ grows with $ \Lambda $ till it reaches its maximun
and then decreases monotonically with $ \Lambda $.  $ d n_s /d \ln k $
vanishes asymptotically for large $\Lambda$ without any oscillation. 
This is a totally {\bf new} feature for hybrid  inflation. 

\medskip

Thus, hybrid inflation describes both $ d n_s /d \ln k >0$ and 
$ d n_s /d \ln k <0$. It must be noticed that $ d n_s /d \ln k < 0$ can reach very low 
values for {\bf small} $\Lambda$ which is a  {\bf totally new} feature 
in hybrid inflation.

\medskip

In summary, the {\bf new} features for the running 
index in hybrid inflation are: both positive and negative running, increasing 
of the running with $\Lambda$, transition from positive to negative running 
passing through zero running when $\Lambda$ grows.
 
\subsection{$r$ vs. $n_s$. Confrontation of Hybrid Inflation to the three
years WMAP data}

\begin{figure}[p]
\begin{turn}{-90}
\centering
\psfrag{"nsr7.dat"}{${\hat \chi}(0)=2.7$}
\psfrag{"nsr10.dat"}{${\hat \chi}(0)=3.3$}
\psfrag{"nsr12.dat"}{${\hat \chi}(0)=3.7$}
\psfrag{"nsr14.dat"}{${\hat \chi}(0)=4.1$}
\psfrag{"nsr16.dat"}{${\hat \chi}(0)=4.5$}
\psfrag{"nsr17.dat"}{${\hat \chi}(0)=4.7$}
\psfrag{"nsr19.dat"}{${\hat \chi}(0)=5.1$}
\psfrag{"nsr20.dat"}{${\hat \chi}(0)=5.3$}
\psfrag{"nsr21.dat"}{${\hat \chi}(0)=5.7$}
\includegraphics[width=10cm,height=14cm]{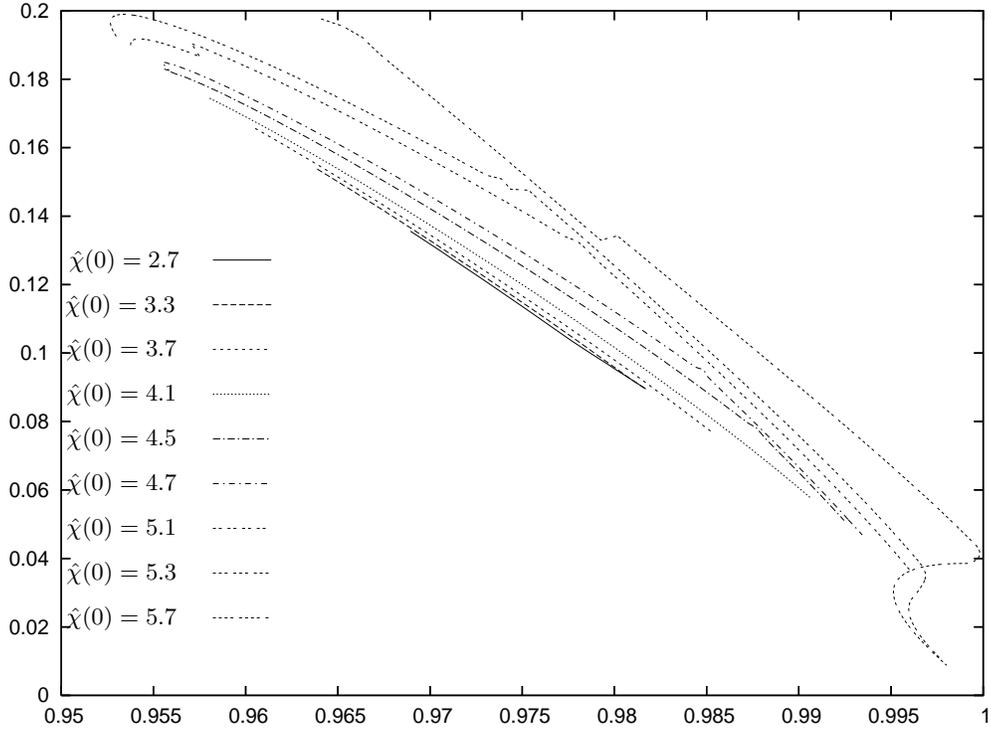}
\end{turn}
\caption{Hybrid inflation. The ratio $ r $ vs. $ n_s $ for $ \mu^2 =
1.7 \; \Lambda > \mu^2_{crit} , \; g^2 = \frac 14 $ and $ 2.7 \; 
\sqrt{\Lambda}\leq \chi(0) \leq 5.7 \; \sqrt{\Lambda} \simeq \chi(0)_{crit} $. 
Notice that here $ n_s < 1 $ for all values of $ \Lambda $ and this range 
of $ \hat  \chi(0) $. We see that $ 0.2 > r > 0.14 $ for the interval
$ 0.952 <  n_s < 0.97$.}
\label{nsrY}
\end{figure}

\begin{figure}[p]
\begin{turn}{-90}
\centering
\psfrag{"nsr21.dat"}{${\hat \chi}(0)=5.7$}
\psfrag{"nsr22.dat"}{${\hat \chi}(0)=6.1$}
\psfrag{"nsr23.dat"}{${\hat \chi}(0)=6.5$}
\psfrag{"nsr24.dat"}{${\hat \chi}(0)=6.9$}
\psfrag{"nsr25.dat"}{${\hat \chi}(0)=7.3$}
\psfrag{"nsr26.dat"}{${\hat \chi}(0)=7.7$}
\psfrag{"nsr27.dat"}{${\hat \chi}(0)=8.1$}
\psfrag{"nsr28.dat"}{${\hat \chi}(0)=8.7$}
\psfrag{"nsr30.dat"}{${\hat \chi}(0)=10.7$}
\psfrag{"nsr33.dat"}{${\hat \chi}(0)=13.7$}
\includegraphics[width=10cm,height=14cm]{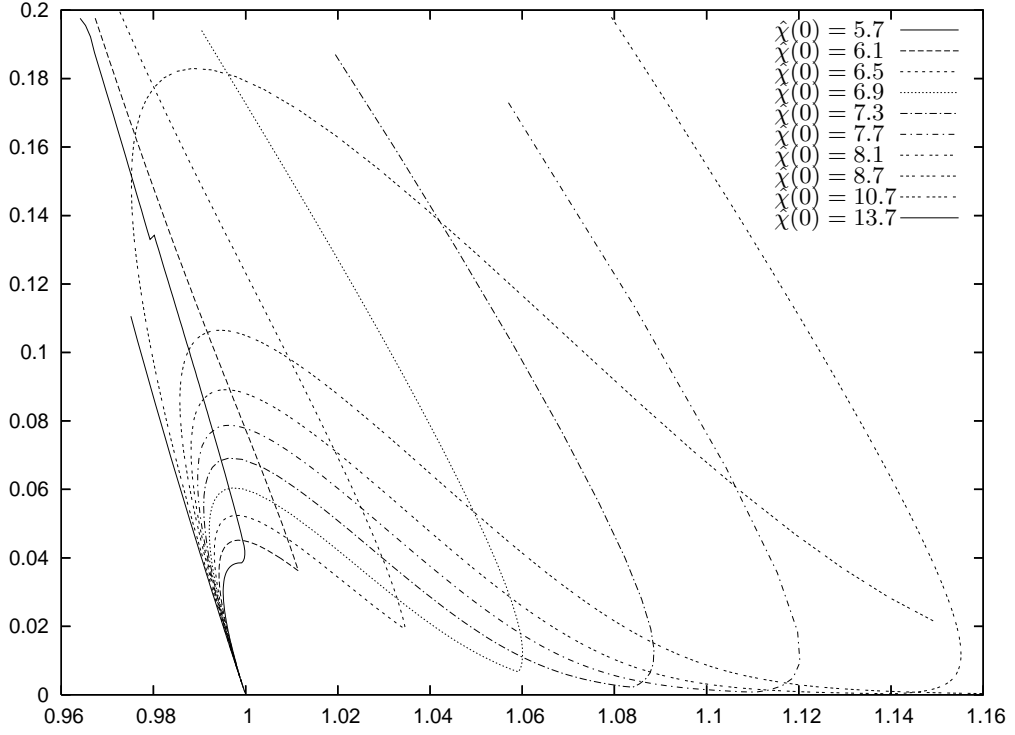}
\end{turn}
\caption{Hybrid inflation. The ratio $ r $ vs. $ n_s $ for $ \mu^2 =
1.7 \; \Lambda  > \mu^2_{crit} , \; g^2 = \frac 14 $ and 
$ \chi(0)_{crit}  \simeq 5.7 \; \sqrt{\Lambda}\leq \chi(0)  \leq 13.7 
\; \sqrt{\Lambda} $. Notice that we have here both $ n_s > 1 $ and  
$ n_s < 1 $ depending on the values of $ \Lambda $ and $ \hat \chi(0) $.
All curves end [$ {\hat \chi(0)} <  {\hat \chi(0)}_{crit} $] or start
 [$ {\hat \chi(0)} > {\hat \chi(0)}_{crit} $] at $ n_s = 1 $ which is the fixed 
point for all values of $ {\hat \chi(0)} $.}
\label{nsrA}
\end{figure}

$r$ vs. $n_s$ in figs. \ref{nsrY}-\ref{nsrA} depicts a oscillatory behavior 
clearly showing the two regimes: ${\hat \chi}(0) > {\hat \chi}(0)_{crit}$ 
corresponding  mainly to $ n_s > 1$, although it also covers a small portion 
of $n_s < 1$, and ${\hat \chi}(0) < {\hat \chi}(0)_{crit}$ for which $n_s$ 
is entirely red tilted. \\

\medskip

All curves end [${\hat \chi}(0) < {\hat \chi}(0)_{crit}$], or start 
[${\hat \chi}(0) > {\hat \chi}(0)_{crit}$], 
at $n_s = 1$, which is the {\bf fixed} point for all ${\hat \chi}(0)$,
with three different behaviors: \\

(1) the sharp decreasing of $r$ in the range $n_s <1$, approaching $n_s =1$ 
as the end point, this is for ${\hat \chi}(0) < {\hat \chi}(0)_{crit}$, 
in which $r$ can take high values for small $n_s$, ($n_s$ near $0.95$). \\

(2) the monotonically decreasing of $r$ with $n_s$ at the right of $r_{max}$, 
for  ${\hat \chi}(0) > {\hat \chi}(0)_{crit}$, in 
which $r$ vanishes asymptotically for `high' $n_s$ , $(n_s > 1.07)$.\\

(3) The {\bf new} hybrid behavior for high ${\hat \chi}(0) > 
{\hat \chi}(0)_{crit}$ in between 
the above two regimes, in which $r$ shows a maximum and a minimum between 
two sharp decreasing and increasing `arms', lying at $n_s < 1$ and $n_s >1$ 
respectively. 
 $r_{min}$ decreases and $r_{max}$ increases as increasing ${\hat \chi}(0) $. 
All $r_{max}$ lie in the red tilted regime $n_s<1$. All $r_{min}$ 
lie in the blue tilted regime $n_s >1$.

\medskip

For ${\hat \chi}(0) > {\hat \chi}(0)_{crit}$, all curves go towards 
$ n_s=1, \; r=0 $. Most of each curve lies in the $ n_s >1 $ 
region, $ r $ monotonically decreases with $n_s$ in the range $ n_s <1 $
from  the maximun $r_{max}$  going  towards $ r = 0 $ for $ n_s \to 1 $.
For $ {\hat \chi}(0) < {\hat \chi}(0)_{crit} $ the curves pill up in the 
$ n_s<1 $ region with almost the same slope, $r$ sharply decreases in this 
region. 

The curves for ${\hat \chi}(0) < {\hat \chi}(0)_{crit}$ pill up in the 
$ n_s<1 $ region with almost the same slope, 
$r$ sharply decreases in this region. 

\begin{figure}[p]
\begin{turn}{-90}
\centering
\psfrag{"nsrun7.dat"}{${\hat \chi}(0)=2.7$}
\psfrag{"nsrun10.dat"}{${\hat \chi}(0)=3.3$}
\psfrag{"nsrun12.dat"}{${\hat \chi}(0)=3.7$}
\psfrag{"nsrun14.dat"}{${\hat \chi}(0)=4.1$}
\psfrag{"nsrun16.dat"}{${\hat \chi}(0)=4.5$}
\psfrag{"nsrun17.dat"}{${\hat \chi}(0)=4.7$}
\psfrag{"nsrun19.dat"}{${\hat \chi}(0)=5.1$}
\psfrag{"nsrun20.dat"}{${\hat \chi}(0)=5.3$}
\psfrag{"nsrun21.dat"}{${\hat \chi}(0)=5.7$}
\includegraphics[width=10cm,height=14cm]{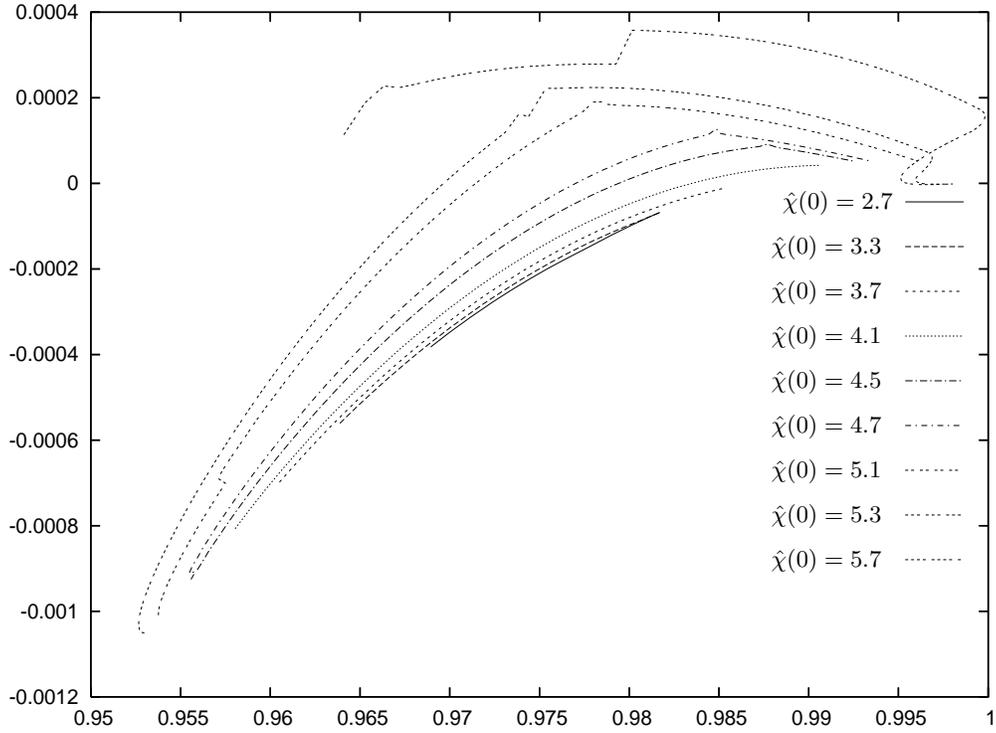}
\end{turn}
\caption{Hybrid inflation. The running $ d n_s/ d \ln k  $ vs. $ n_s $ for 
$ \mu^2 = 1.7 \; \Lambda  > \mu^2_{crit}, \; g^2 = \frac 14 $ and 
$ 1.9 \; \sqrt{\Lambda}\leq \chi(0)  \leq 5.7 \; \sqrt{\Lambda} \simeq 
\chi(0)_{crit} $. Notice that we have here both positive and negative running
$  d n_s/ d \ln k $ depending on the values of $ \Lambda $ and $ \hat \chi(0) $.
Most values of the running are negative in this regime. For $ 0.952 <  n_s < 0.97 $,
we have $ -0.001 < d n_s /d \ln k < 0 $.} \label{nsrunY}
\end{figure}

\begin{figure}[p]
\begin{turn}{-90}
\centering
\psfrag{"nsrun21.dat"}{${\hat \chi}(0)=5.7$}
\psfrag{"nsrun22.dat"}{${\hat \chi}(0)=6.1$}
\psfrag{"nsrun23.dat"}{${\hat \chi}(0)=6.5$}
\psfrag{"nsrun24.dat"}{${\hat \chi}(0)=6.9$}
\psfrag{"nsrun25.dat"}{${\hat \chi}(0)=7.3$}
\psfrag{"nsrun26.dat"}{${\hat \chi}(0)=7.7$}
\psfrag{"nsrun27.dat"}{${\hat \chi}(0)=8.1$}
\psfrag{"nsrun28.dat"}{${\hat \chi}(0)=8.7$}
\psfrag{"nsrun30.dat"}{${\hat \chi}(0)=10.7$}
\psfrag{"nsrun33.dat"}{${\hat \chi}(0)=13.7$}
\includegraphics[width=10cm,height=14cm]{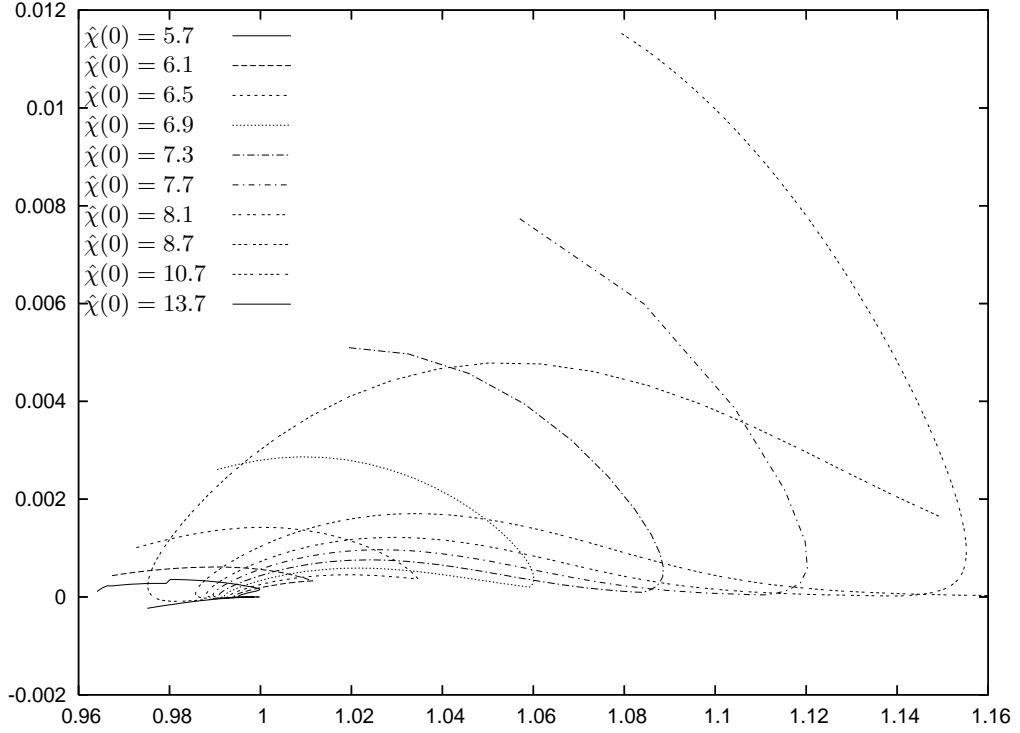}
\end{turn}
\caption{Hybrid inflation. The running $  d n_s/ d \ln k $ vs. $ n_s $ 
for $ \mu^2 = 1.7 \; \Lambda  > \mu^2_{crit}, \; g^2 = \frac 14 $ 
and $ \chi(0)_{crit}  \simeq 5.7 \; \sqrt{\Lambda} \leq \chi(0)  
\leq 13.7 \; \sqrt{\Lambda} $. Notice that we have here both positive and 
negative running $  d n_s/ d \ln k $ depending on the values of 
$ \Lambda $ and $ \hat \chi(0) $.}
\label{nsrunA}
\end{figure}

\begin{figure}[p]
\begin{turn}{-90}
\centering
\psfrag{"nsr93h20"}{border for new inflation}
\psfrag{"nsr93h0"}{border for new inflation}
\psfrag{"nsr33"}{border for hybrid inflation}
\includegraphics[width=10cm,height=14cm]{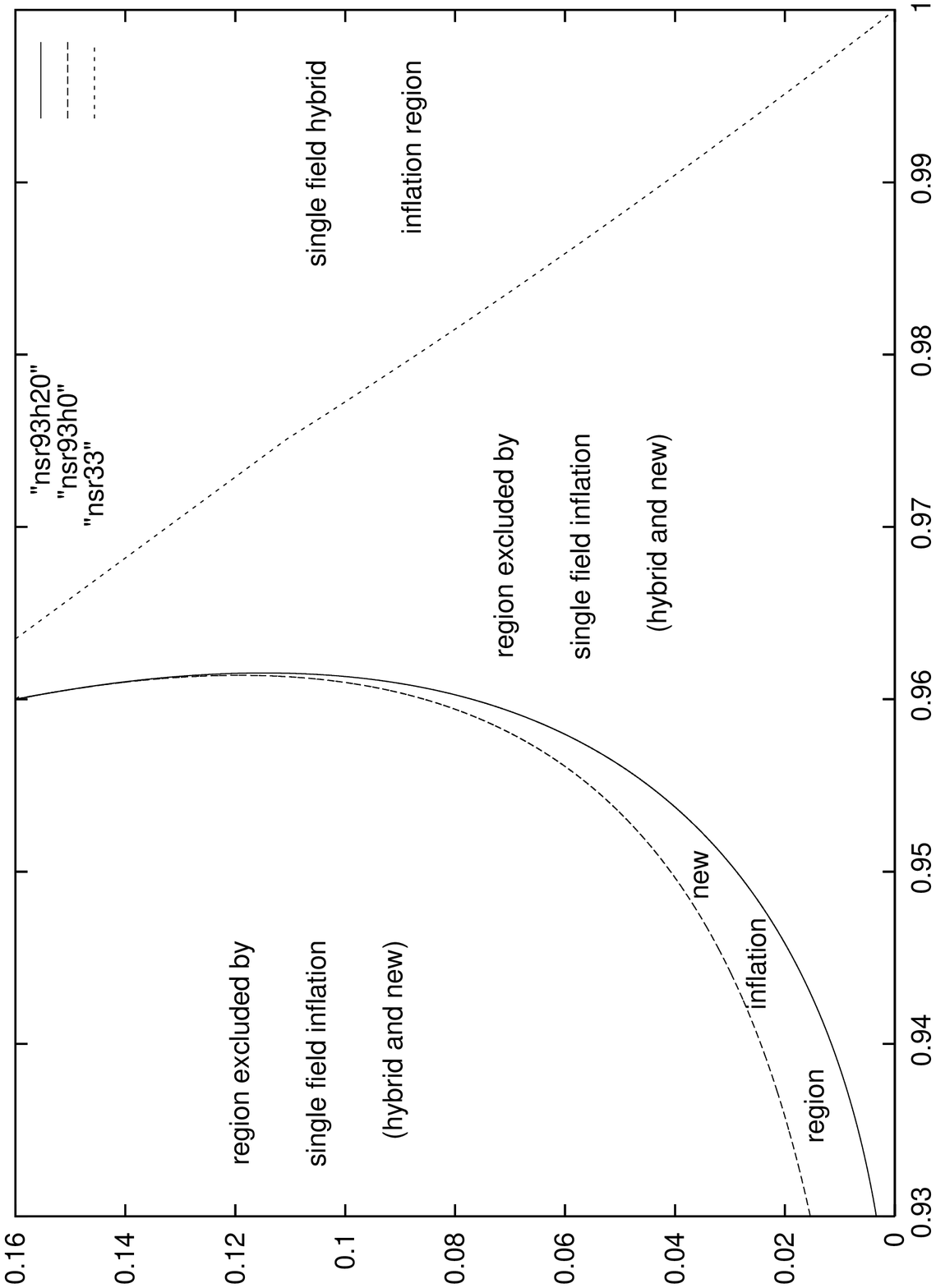}
\end{turn}
\caption{Regions described in the $ (r, n_s)$-plane by new and single field
hybrid inflation for $ n_s < 1 $. The hybrid inflation border corresponds to 
$ \mu^2 = 1.7 \; \Lambda $. For $ n_s > 1 $, all values of $ (r, n_s) $ 
can be described by hybrid inflation (at least for $ r < 0.2, \; n_s < 1.15 $).
The excluded region cannot be described by single field inflation 
(neither hybrid inflation, nor new inflation). Two or more fields inflation
could describe such regions.}
\label{bordenuevo}
\end{figure}

All the blue tilted values of $ (n_s, r) $ in the domain $ 1 < n_s < 1.15 , 
\; 0 < r < 0.2 $ are realized by hybrid inflation. 

\medskip

The red tilted regime in hybrid inflation can only be realized
for $  \mu^2 >  \mu^2_{crit} \simeq 0.13 \; \Lambda $.
Moreover, the possible values of  $ (n_s, r) $ are in the upper-right quadrant 
as shown in fig. \ref{bordenuevo}.

\medskip

We see at the light of the three years WMAP data ref. \cite{WMAP3}
that hybrid inflation in the blue tilted regime $ n_s > 1 $ is ruled out 
[eq.(\ref{nswmap})] \cite{WMAP3}. That is, hybrid inflation in the regime 
$ {\hat\mu^2} < {\hat\mu^2_{crit}} $ is ruled out as well as hybrid inflation
in the regime $ {\hat\mu^2} >  {\hat\mu^2_{crit}} $ with 
$ {\hat \chi}(0) > {\hat \chi}(0)_{crit} $.

\medskip

Hybrid inflation in the red tilted regime $ {\hat \mu^2} > {\hat\mu^2_{crit}} $
and $ {\hat \chi}(0) < {\hat \chi}(0)_{crit} $
fulfills the three years WMAP value for $ n_s $ eq.(\ref{nswmap}), as
well as the bound on the ratio $ r $ eq.(\ref{rwmap}).
We can read from fig. \ref{nsrY} that 
$$ 
0.2 > r < 0.14  \quad {\rm for} \quad  0.952 <  n_s < 0.97 \; .
$$
In addition, we find in fig. \ref{nsrunY}
negative values for the running in this range, that is:
$$ 
-0.001 < d n_s /d \ln k < 0 \quad {\rm for} \quad 0.952 <  n_s < 0.97 \; .
$$
There are clearly two regions which are not covered by hybrid inflation 
with only one inflaton field, neither by new inflation as shown 
in fig. \ref{bordenuevo}.

\bigskip

Simple single field  inflation models have been recently studied within a 
numerical approach \cite{BST}. In ref. \cite{PE} cosmological
data are fitted with the help of a Markov Chain Monte Carlo analysis.
 
\medskip

There is an interplay between the bounds of neutrino masses and the sign
of $ (n_s - 1) $. A non-zero neutrino mass decreases the power in the 
small scales (large wavenumbers $k$). The same happens if $ n_s $ becomes 
smaller than unit.
Therefore, {\it if} $  n_s < 1 $, the power spectrum permits more stringent
tests of the neutrino masses \cite{fuku}. The effect of neutrino masses
for small scales $ n_s > 1 $ can be cancelled by a spectral index $ n_s > 1 $.

\bigskip

{\bf Acknowledgment:} we thank Daniel Boyanovsky for useful discussions.


\begin{thebibliography}{99}
\bibitem{guth} D. Kazanas, ApJ 241, L59 (1980);
A. Guth, Phys. Rev. \textbf{D23}, 347 (1981);
K. Sato, MNRAS, {\bf 195}, 467 (1981).

\bibitem{mukyotr} V. F. Mukhanov , G. V. Chibisov, Soviet Phys.
JETP Lett. \textbf{33}, 532 (1981). 
S. W. Hawking, Phys. Lett. \textbf{B115}, 295
(1982). A. H. Guth , S. Y. Pi, Phys. Rev. Lett.
\textbf{49}, 1110 (1982). A. A. Starobinsky, Phys. Lett.
\textbf{B117}, 175 (1982). J. M. Bardeen, P. J. Steinhardt , M. S.
Turner, Phys. Rev. \textbf{D28}, 679 (1983). 
V. F. Mukhanov, H. A. Feldman , R. H.
Brandenberger, Phys. Rept. \textbf{215}, 203 (1992).

\bibitem{libros}
P. Coles and F. Lucchin, {\em Cosmology}, John Wiley,
Chichester, 1995. A. R. Liddle and D. H. Lyth,
{\em Cosmological Inflation and Large Scale Structure}, Cambridge University
Press, 2000. S. Dodelson, {\em Modern Cosmology}, Academic Press, 2003.
D. H. Lyth , A. Riotto, Phys. Rept. \textbf{314}, 1 (1999).
\bibitem{WMAP} C. L. Bennett \emph{et.al.} (WMAP collaboration),
Ap. J. Suppl. \textbf{148}, 1 (2003).

A. Kogut  \emph{et.al.} (WMAP collaboration),
Ap. J. Suppl. \textbf{148}, 161 (2003).

D. N. Spergel \emph{et. al. }(WMAP collaboration),
Ap. J. Suppl. \textbf{148}, 175 (2003).

H. V. Peiris \emph{et.al.} (WMAP collaboration), Ap. J.
Suppl.\textbf{148}, 213 (2003).

\bibitem{WMAP3}
D. N. Spergel \emph{et. al. }(WMAP collaboration),
astro-ph/0603449.

 L. Page, \emph{et. al. }(WMAP collaboration),
astro-ph/0603450.

G. Hinshaw,\emph{et. al. }(WMAP collaboration),
astro-ph/0603451.

 N. Jarosik, \emph{et. al. }(WMAP collaboration),
astro-ph/0603452.

\bibitem{2dF} A. G. S\'anchez et al., Mon. Not. Roy. Astron. Soc. {\bf 366}, 
189 (2006).

\bibitem{SDSS} U. Seljak et al.,  Phys. Rev. D71, 103515 (2005).

\bibitem{Teg} M. Tegmark wt al.,  Phys. Rev. D69, 103501 (2004).

\bibitem{1sN} D. Boyanovsky, H. J. de Vega, N. G. S\'anchez,
Phys. Rev. {\bf D 73}, 023008 (2006).

\bibitem{nos} D. Cirigliano,  H. J. de Vega, N. G. Sanchez,
Phys. Rev. {\bf D 71}, 103518 (2005).

\bibitem{quir} L. D. Landau, E. M. Lifshits, `Physique Statistique',
chapter 14, Mir Ellipses, Paris, 1994.
H. Leutwyler, Ann. Phys. 235, 165 (1994), hep-ph/9409423.
S. Weinberg, hep-ph/9412326 and `The Quantum Theory of Fields', vol.
2, Cambridge University Press, Cambridge, 2000.

\bibitem{BST} L. A. Boyle, P. J. Steinhardt, N. Turok, Phys. Rev. Lett., 
{\bf 96}, 111301, (2006).

\bibitem{PE} H. V. Peiris, R. Easther, astro-ph/0603587.

\bibitem{fuku} M. Fukugita, hep-ph/0511068.

\bibitem{barrow} A. R. Liddle, P. Parsons , J. D. Barrow, Phys.
Rev. \textbf{D50}, 7222 (1994).

\bibitem{hu} See for example,  W. Hu and  S. Dodelson,
Ann. Rev. Astron. Ap. \textbf{40}, 171 (2002); J. Lidsey, A.
Liddle, E. Kolb, E. Copeland, T. Barreiro and M. Abney, Rev. of
Mod. Phys. {\bf 69}, 373, (1997). W. Hu, astro-ph/0402060.

\bibitem{lin} A. Linde,  Phys. Rev. D49, 748 (1994).
J. Garc\'{\i}a Bellido,  A. Linde,  Phys. Rev. D57, 6075 (1998).
E. J. Copeland, A. R. Liddle, D. H. Lyth, E. D. Stewart, 
D. Wands, Phys. Rev. D49, 6410 (1994). 
\end{thebibliography}
\end{document}